\documentclass[12pt,preprint]{aastex}
\usepackage{natbib}
\usepackage{epsfig,latexsym,graphicx,epsf,subfigure}
\usepackage{threeparttable,hhline,longtable}
\newcommand{\om}{\Omega_{\rm M}}
\newcommand{\ola}{\Omega_\Lambda}

\newcommand{\chisq}{\chi^2}
\newcommand{\eqref}[1]{(\ref{#1})}

\shortauthors{Nobili et al.}
\begin{document}

\title{Constraining dust and color variations of high-z SNe using NICMOS on
the Hubble Space Telescope\footnote{Based on observations made with the NASA/ESA Hubble Space
  Telescope, obtained from the data archive at the Space Telescope
  Institute. STScI is operated by the association of Universities for
  Research in Astronomy, Inc. under the NASA contract NAS 5-26555. The
  observations are associated with program GO-07850.}}   

\author{S.~Nobili\altaffilmark{1},
V.~Fadeyev\altaffilmark{2},
G.~Aldering\altaffilmark{3},
R.~Amanullah\altaffilmark{1,3,4,5},
K. ~Barbary\altaffilmark{3,6},
M.~S.~Burns\altaffilmark{7},
K.~S.~Dawson\altaffilmark{3},
S.~E.~Deustua\altaffilmark{8},
L.~Faccioli\altaffilmark{3,4},
A.~S.~Fruchter\altaffilmark{8},
G.~Goldhaber\altaffilmark{3,6},
A.~Goobar\altaffilmark{1,5},
I.~Hook\altaffilmark{9},
D.~A.~Howell\altaffilmark{10},
A.~G.~Kim\altaffilmark{3},
R.~A.~Knop\altaffilmark{11},
C.~Lidman\altaffilmark{5,12},
J.~Meyers\altaffilmark{3,6},
P.~E.~Nugent\altaffilmark{3},
R.~Pain\altaffilmark{13},
N.~Panagia\altaffilmark{8},
S.~Perlmutter\altaffilmark{3,6},
D.~Rubin\altaffilmark{3,6},
A.~L.~Spadafora\altaffilmark{3},
M. ~Strovink\altaffilmark{3,6},
N.~Suzuki\altaffilmark{3}, and
H.~Swift\altaffilmark{3,6}
\newline
\newline
(The Supernova Cosmology Project)}
\email{serena@physto.se}

\altaffiltext{1}{Department of Physics, Stockholm University,
  Albanova University Center, S-106 91 Stockholm, Sweden} 
 
\altaffiltext{2}{Santa Cruz Institute for Particle Physics, University of California, Santa Cruz,
CA 95064, USA} 
\altaffiltext{3}{E. O. Lawrence Berkeley National Laboratory, 1
  Cyclotron Rd., Berkeley, CA 94720, USA } 
\altaffiltext{4}{Space Sciences Laboratory, University of California
  Berkeley, Berkeley, CA 94720, USA} 
\altaffiltext{5}{The Oskar Klein Center, Stockholm University,
S-106 91 Stockholm, Sweden}
\altaffiltext{6}{Department of Physics, University of California
  Berkeley, Berkeley, 94720-7300 CA, USA} 
\altaffiltext{7}{Colorado College, 14 East Cache La Poudre St.,
  Colorado Springs, CO 80903, USA} 
\altaffiltext{8}{Space Telescope Science Institute, 3700 San Martin
  Drive, Baltimore, MD 21218, USA} 
\altaffiltext{9}{Sub-Department of Astrophysics, University of Oxford,
  Denys Wilkinson Building, Keble Road, Oxford OX1 3RH, UK} 
\altaffiltext{10}{Department of Astronomy and Astrophysics, University
  of Toronto, 60 St. George St., Toronto, Ontario M5S 3H8, Canada} 
\altaffiltext{11}{Meta-Institute of Computational Astrophysics,
  www.mica-vw.org} 
\altaffiltext{12}{European Southern Observatory, Alonso de Cordova
  3107, Vitacura, Casilla 19001, Santiago 19, Chile } 
\altaffiltext{13}{LPNHE, CNRS-IN2P3, University of Paris VI \& VII,
  Paris, France}

\date{Received ...; accepted ...}

\date{} \begin{abstract}
  
  We present data from the Supernova Cosmology Project for five high
  redshift Type Ia supernovae (SNe~Ia) that were obtained using the
  NICMOS infrared camera on the Hubble Space Telescope.  We add two
  SNe from this sample to a rest-frame $I$-band Hubble diagram,
  doubling the number of high redshift supernovae on this diagram.
  This $I$-band Hubble diagram is consistent with a flat universe (
  $\om,\;\ola$) = (0.29, 0.71). A homogeneous distribution of large
  grain dust in the intergalactic medium (replenishing dust) is
  incompatible with the data and is excluded at the 5$\sigma$
  confidence level, if the SN host galaxy reddening is corrected
  assuming $R_V=1.75$.  We use both optical and infrared observations
  to compare photometric properties of distant SNe~Ia with those of
  nearby objects. We find generally good agreement with the expected
  color evolution for all SNe except the highest redshift SN in our
  sample (SN~1997ek at $z= 0.863$) which shows a peculiar color
  behavior. We also present spectra obtained from ground based
  telescopes for type identification and determination of redshift.
\end{abstract}

\section{Introduction}
\label{sec:intro}

A decade ago, the development of Type Ia supernova (SNe~Ia) as
distance indicators began making possible precise measurements of the
expansion history, leading to the understanding that a dark energy
dominates the fate of our Universe
\citep{1998Natur.391...51P,1998ApJ...509...74G,1998ApJ...507...46S,1998AJ....116.1009R,1999ApJ...517..565P}.
With the advent of increasingly larger and more precise SNe Ia
surveys, which aim to elucidate the nature of dark energy, it has
become clear that systematic uncertainties are a limiting factor
\citep{2003ApJ...598..102K,2003ApJ...594....1T,2006A&A...447...31A,2007ApJ...666..694W,2007ApJ...659...98R,2008ApJ...686..749K}.
In particular, knowledge of SNe intrinsic colors and dimming due to
extinction by dust in the host galaxy have been identified as the
dominant systematic effects for the determination of cosmological
parameters, see e.g., \citet{2008JCAP...02..008N}.  It is often
assumed that dust in distant galaxies resembles the dust in the Milky
Way, whose properties are well studied and for which maps of dust
distribution and composition are available. This question, however,
remains unresolved.  Another source of uncertainty is in our knowledge
of the SN intrinsic color and its dispersion, which is normally used
for determining the amount of reddening\footnote{One normally refers
  to reddening as an effect of dust extinction, either in the SN host
  galaxy or in the Milky-Way. However, there could be a relation
  between colors and intrinsic brightness of SNe~Ia that mimics an
  extinction law, though not caused by differential dimming by
  dust. Thus, we use the term color excess to describe the deviation
  from the average color, irrespective of the underlying mechanism
  causing the offset.} for each supernova.  Moreover, the
understanding of the mechanisms leading to the explosion of SNe~Ia or
the nature of the progenitor systems is still incomplete. Thus, the
empirical properties of the standard candles need to be continuously
compared between distant and nearby objects.  While there is evidence
for evolution in average SN properties with redshift
\citep{2007ApJ...667L..37H}, no intrinsic SN evolution has been
observed in the spectral and photometric properties of individual
objects (see e.g.,
\citet{2005A&A...437..789N,2006AJ....131.1648B,2007A&A...470..411G,2008ApJ...674...51E,
  2008ApJ...684...68F,2008A&A...477..717B,2007ApJ...659...98R}).

Dust extinction has a strong wavelength dependence. 
Milky-Way type dust causes 
twice as much extinction in the $B$-band as in the $I$-band, and
a factor of 4 more in the $B$-band than in the $J$-band. This suggests
that observations in the near-infrared and infrared ($\lambda > 0.7\
\mu m$) are a very powerful tool 
for minimizing systematic uncertainties in supernova
cosmology due to the determination of the host galaxy extinction
\citep{2005A&A...437..789N,2008ApJ...689..377W}. 
However, ground-based infrared observations require relatively long 
exposures and are limited by atmospheric emission and absorption.
Observations from space have significantly better resolution and less
background contamination leading to more precise infrared photometry
from space than from the ground.

In this paper, we use Near-Infrared Camera and Multi-Object
Spectrometer (NICMOS) infrared and Wide Field Planetary Camera 2 (WFPC2)
optical data to study the colors of five distant SNe~Ia over a broad
range of rest-frame wavelengths. In Section \ref{sec:data}, we present
the photometric data. In Section \ref{sec:spectra}, we show the
spectra collected for the type identification and determination of
redshift. We compare the observed optical and infrared colors with
those of nearby SNe~Ia in Section \ref{sec:colour}. In Section
\ref{sec:HD}, we add two SNe from our sample to the current rest-frame
$I$-band Hubble diagram, doubling the number of $z>0.1$ SNe and
extending the analysis to $z=0.638$. Finally, we discuss our results
in Section \ref{sec:discussion}.

\section{The data set}
\label{sec:data}

We present data of 5 distant Type Ia supernovae that were obtained
with the NICMOS mounted on the Hubble Space Telescope (HST). These
supernovae, in the redshift range $0.356<z<0.862$, were discovered by
the Supernova Cosmology Project during ground based searches for
SNe~Ia conducted with the 4m Blanco telescope at the Cerro Tololo
Inter-American Observatory (CTIO) in 1997 and 1998. Optical follow up
observations were carried out from the ground and from space and have
been presented in previous work \citep{2003ApJ...598..102K}. These
consist of multi-epoch observations with the WFPC2 camera on HST in
the $F675W$ and $F814W$ broadband filters and multi-epoch ground-based
observations in the $R$ and $I$ bands.  The HST filter pair and the
ground-based filter pair both correspond to rest-frame $B$ and $V$
for $z < 0.7$ and to rest-frame $U$ and $B$ for $z > 0.7$.

\begin{figure}
  \includegraphics[width=8cm]{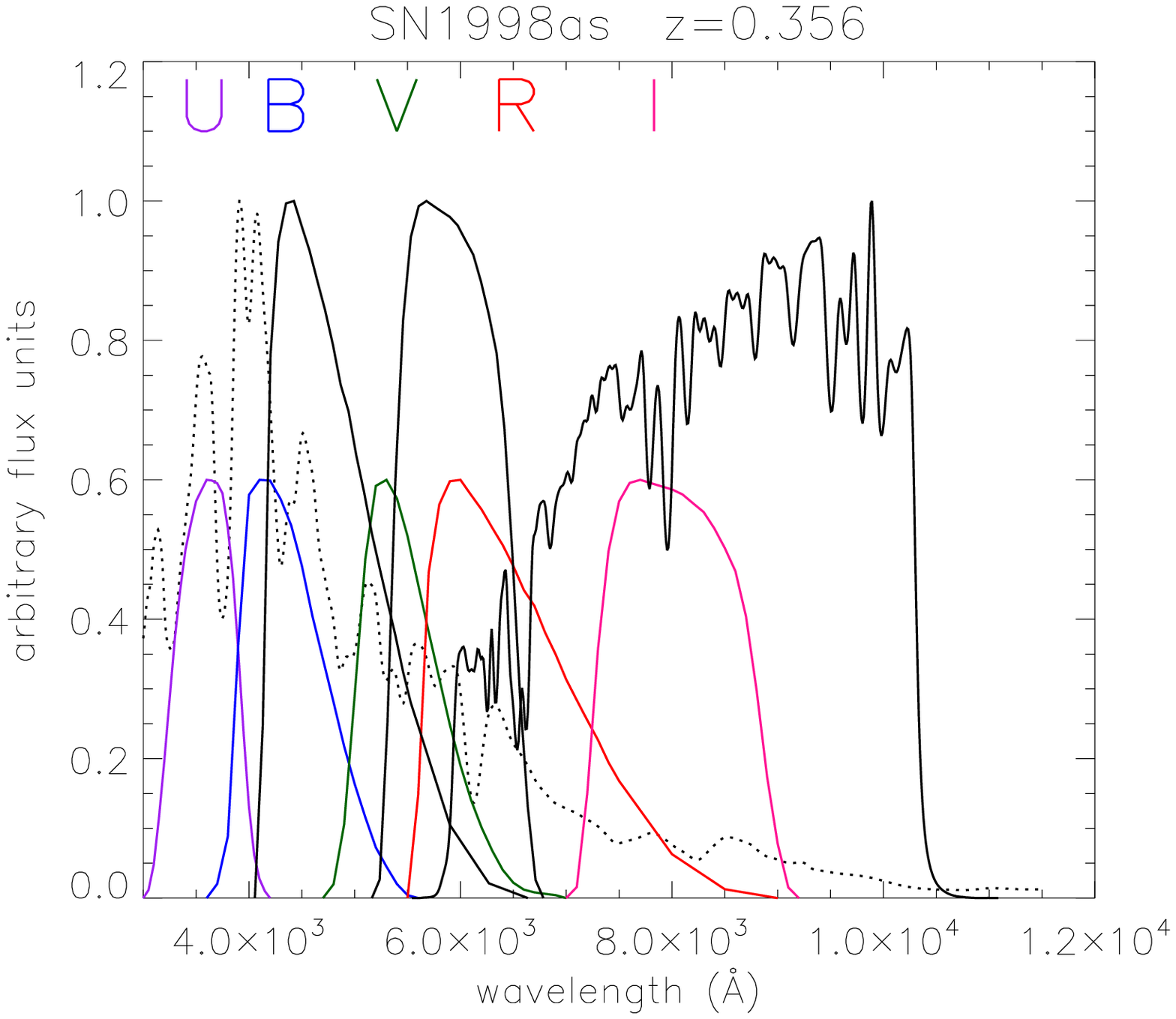}
  \includegraphics[width=8cm]{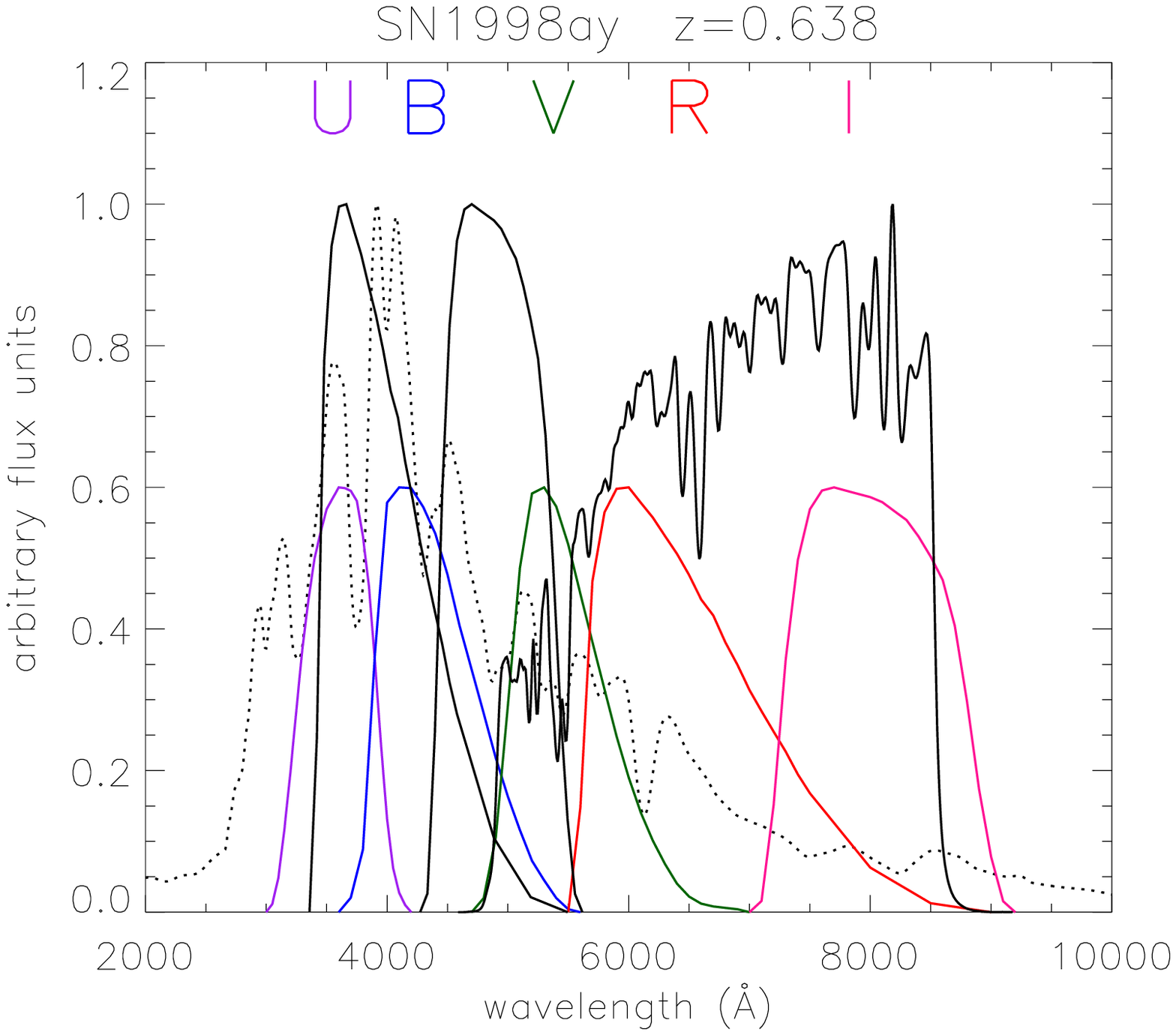}
    \includegraphics[width=8cm]{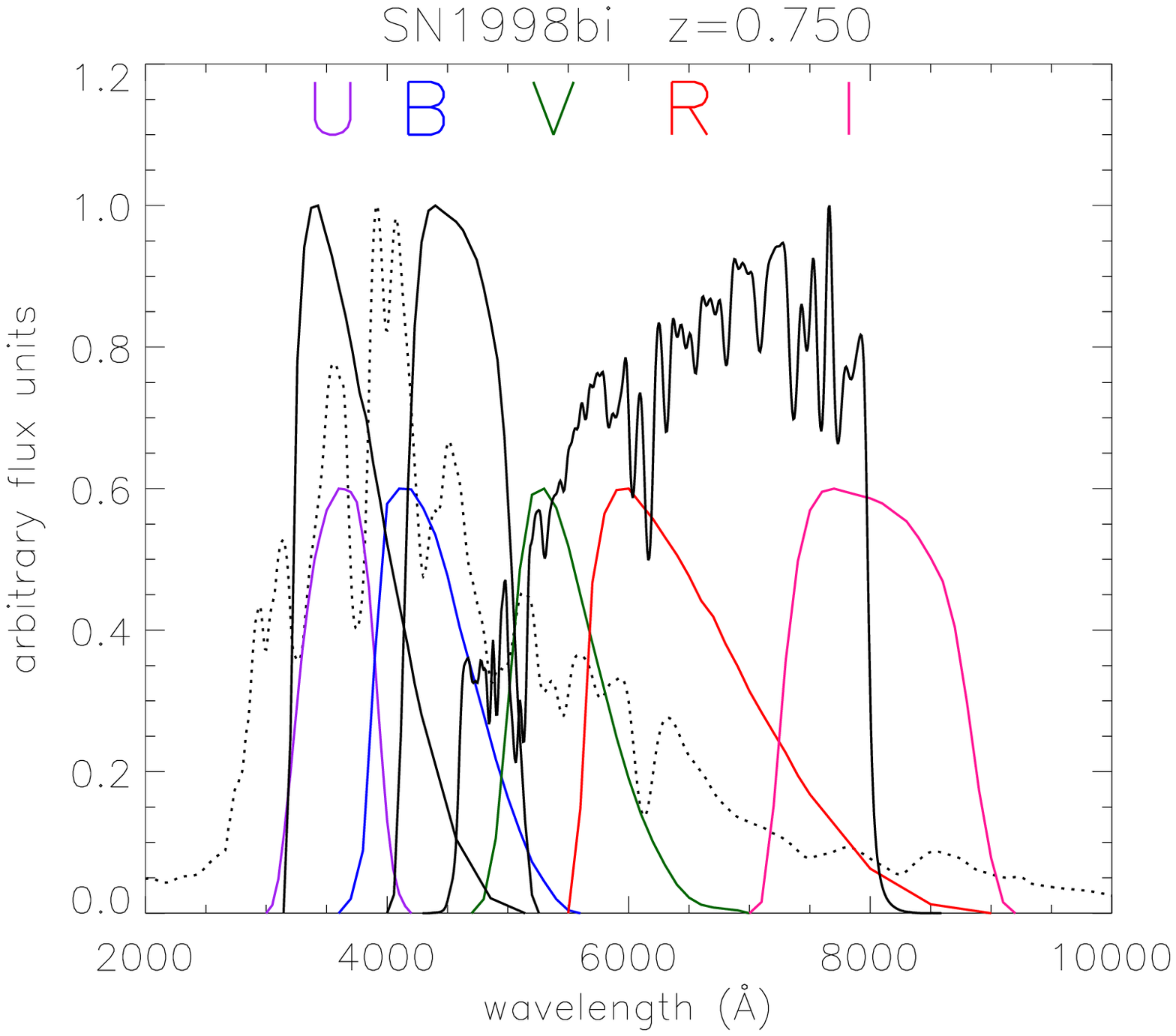}  
\includegraphics[width=8cm]{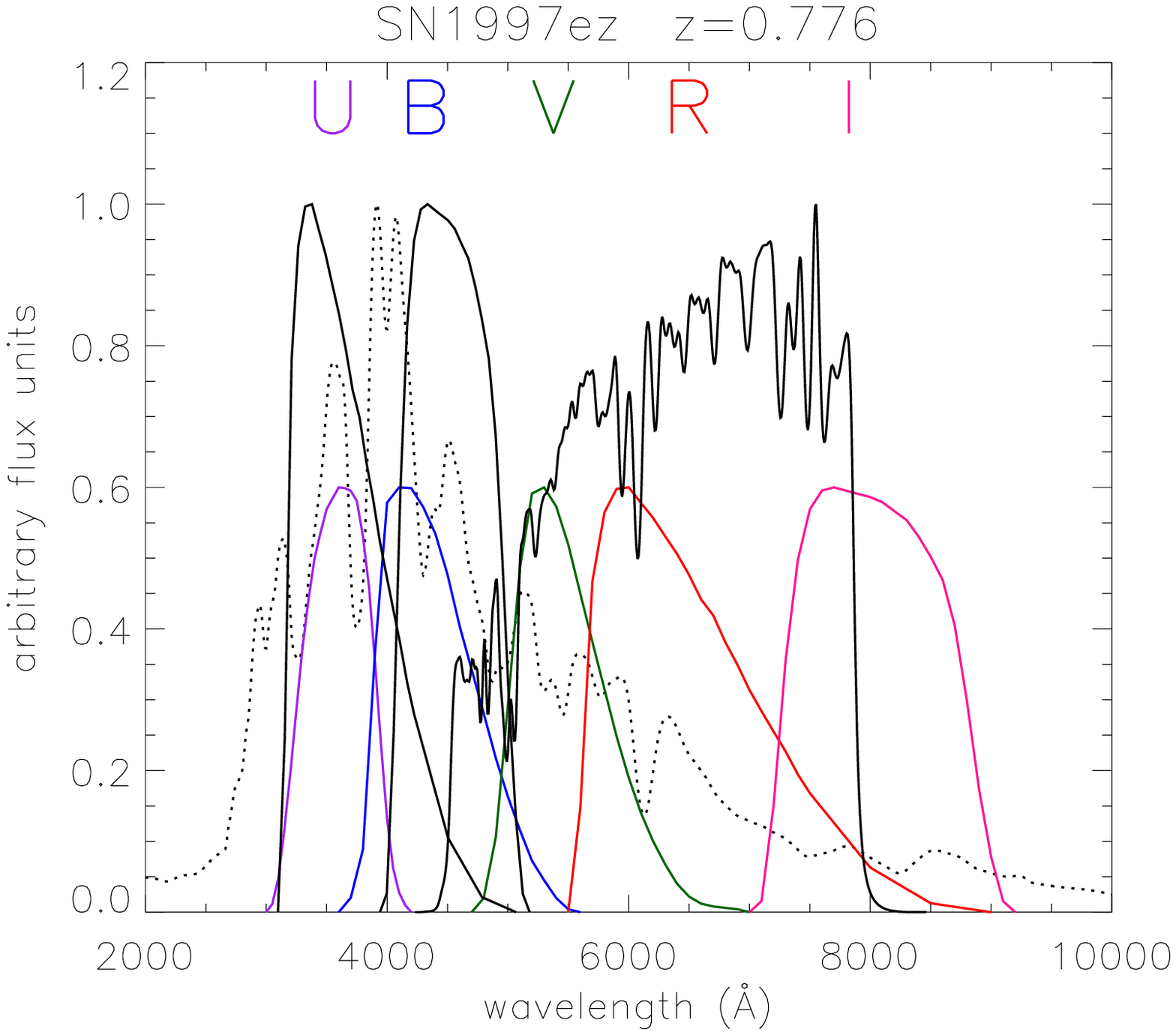}
  \includegraphics[width=8cm]{{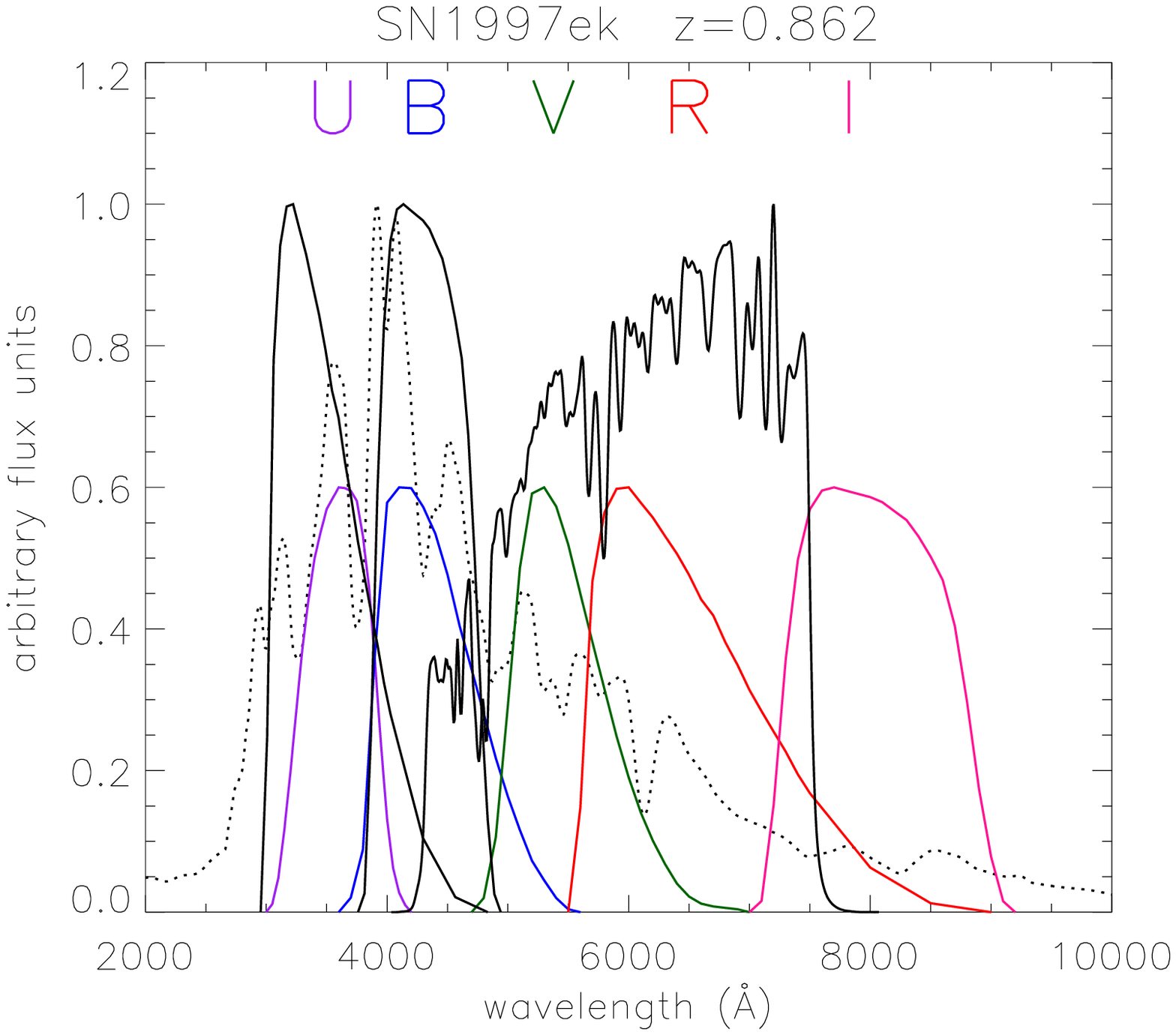}}
\caption{Correspondence of the observed $R$, $I$ and $F110W$ passbands
  (normalized to 1) de-reshifted to  match the rest-frame optical
  filters (arbitrarily normalized to 0.6). The dotted 
  line show the SNe~Ia spectral template at the time of maximum.} 
\label{filters}
\end{figure}

Infrared images were obtained using the broadband $F110W$ filter
on the NIC2 camera. Given the broad redshift range of the SNe,
the $F110W$ filter corresponds to different rest-frame filters as shown in 
Figure \ref{filters}.

Some of the NICMOS data were affected by the passage of HST through the
South Atlantic Anomaly prior to the observations.  After visual
examination, good images were processed with a modified version of the
STScI pipeline. Details of the modifications are explained in
\citet{2006PASP..118..907F} and include an improved technique for
evaluating per-pixel count rate errors and a different method for
correcting the cosmic ray (CR) hits. Although the standard STScI
pipeline also evaluates the count rate errors, the modifications
improved the statistical correspondence between the errors and
fluctuations in the sky level and were used to estimate the
photometric errors.  The standard bias equalization procedure was used
during the processing. The images were corrected for the rate
dependent non-linearity effect \citep{2005ISR02,2006ISR03}.

PSF photometry was done on the processed images
with TinyTIM models \citep{2004krist.book}. We have used standard star
observations to verify that PSF photometry with
TinyTIM is consistent with simple aperture photometry to 1\% accuracy.
The procedure was similar to the chi-square fitter used for optical photometry
in \citet{2003ApJ...598..102K}. 
There were two main differences due to features of the infrared
data, as compared to the CCD-based observations. (1) The pixel
flux uncertainty was taken from the modified pipeline evaluation.
(2) Bad pixels, a frequent occurrence in NICMOS arrays, were
explicitly omitted from the chi-square evaluation.

For most SNe, the distance to the host galaxy was sufficiently large
that the host galaxy light could be modeled by a second
degree polynomial. In these cases, the reference images were
used to visually examine the SN location for unusual host galaxy
shapes. The only exception was SN~1998ay, where an elliptical-like
galaxy model was used in a joint fit with the final reference images.
We have evaluated the systematic uncertainties due to host galaxy
modeling and CR rejection to be typically at the 1--4\% level,
although there were a couple of cases as large as 10--18\%.

Because the NICMOS detectors have a count-rate dependent non-linearity
we correct for (often referred to as the Bohlin Effect \citep{2008ISR003}),
we must apply a corresponding zeropoint, $Z_{non-lin}$, to derive Vega
magnitudes. Our data were acquired when the NICMOS IR arrays were
operating at a temperature of 60 K. For this temperature, STScI
only provides zero points for standard calibrations.  The well modeled
detector non-linearity corrections are based on an operating temperature of
77 K.  We assume that the non-linear corrections at 60 K are the same
as at 77 K\footnote{This is supported
by the fact that the NICMOS array
response only changed by about 50\%\ after installation of the cryocooler.
On the scale of the non-linearity correction, 0.05 mag per decade of flux
change, the effect of extra 50\%\ flux nominally results in 0.009 mag correction,
a negligible amount.}, namely, $Z_{std}^{60K}- Z_{non-lin}^{60K}
= Z_{std}^{77K}- Z_{non-lin}^{77K} = 0.179 mag$. We obtain
$Z_{non-lin}^{60K}$ = 22.268. 

\begin{figure}
 \includegraphics[width=7cm]{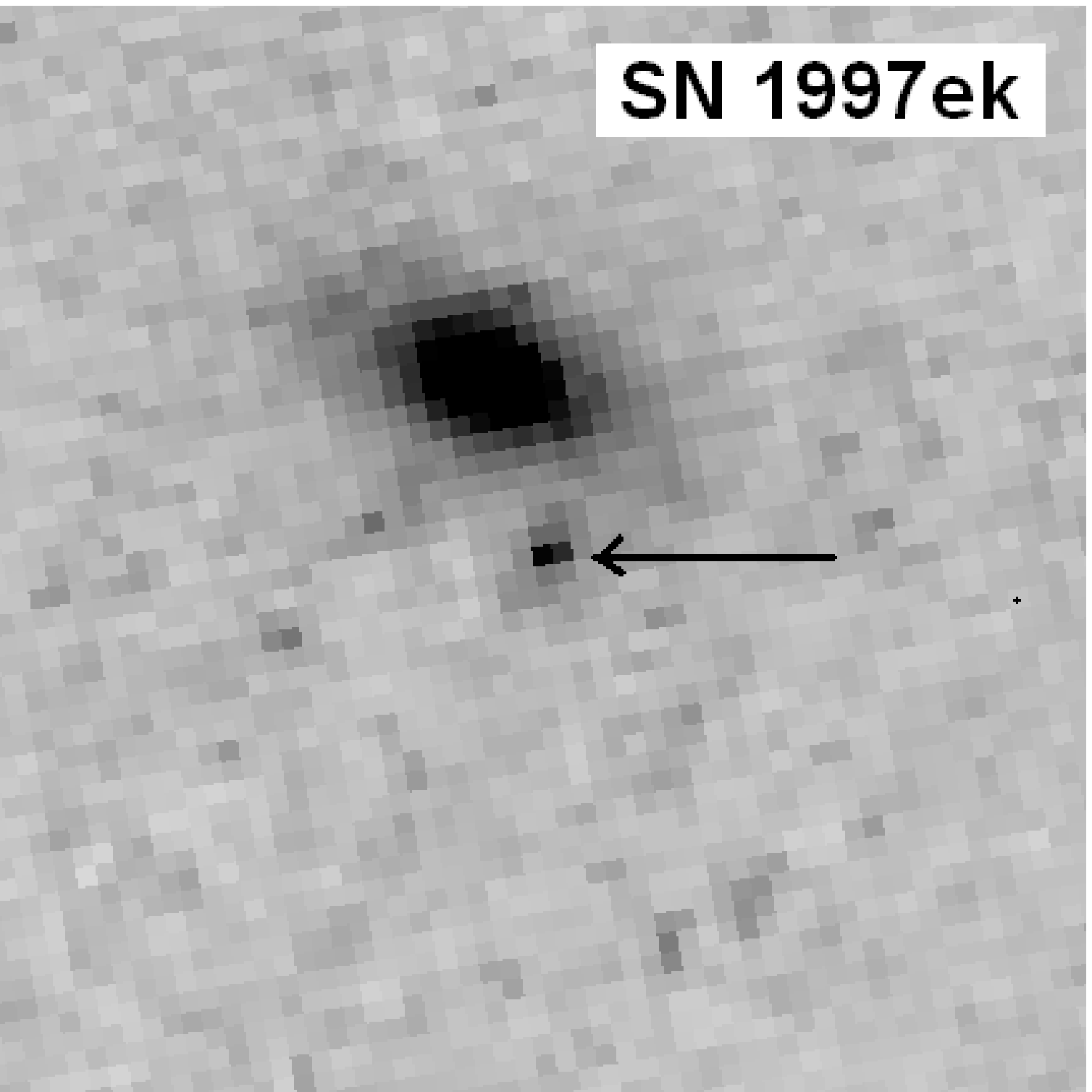}
 \includegraphics[width=7cm]{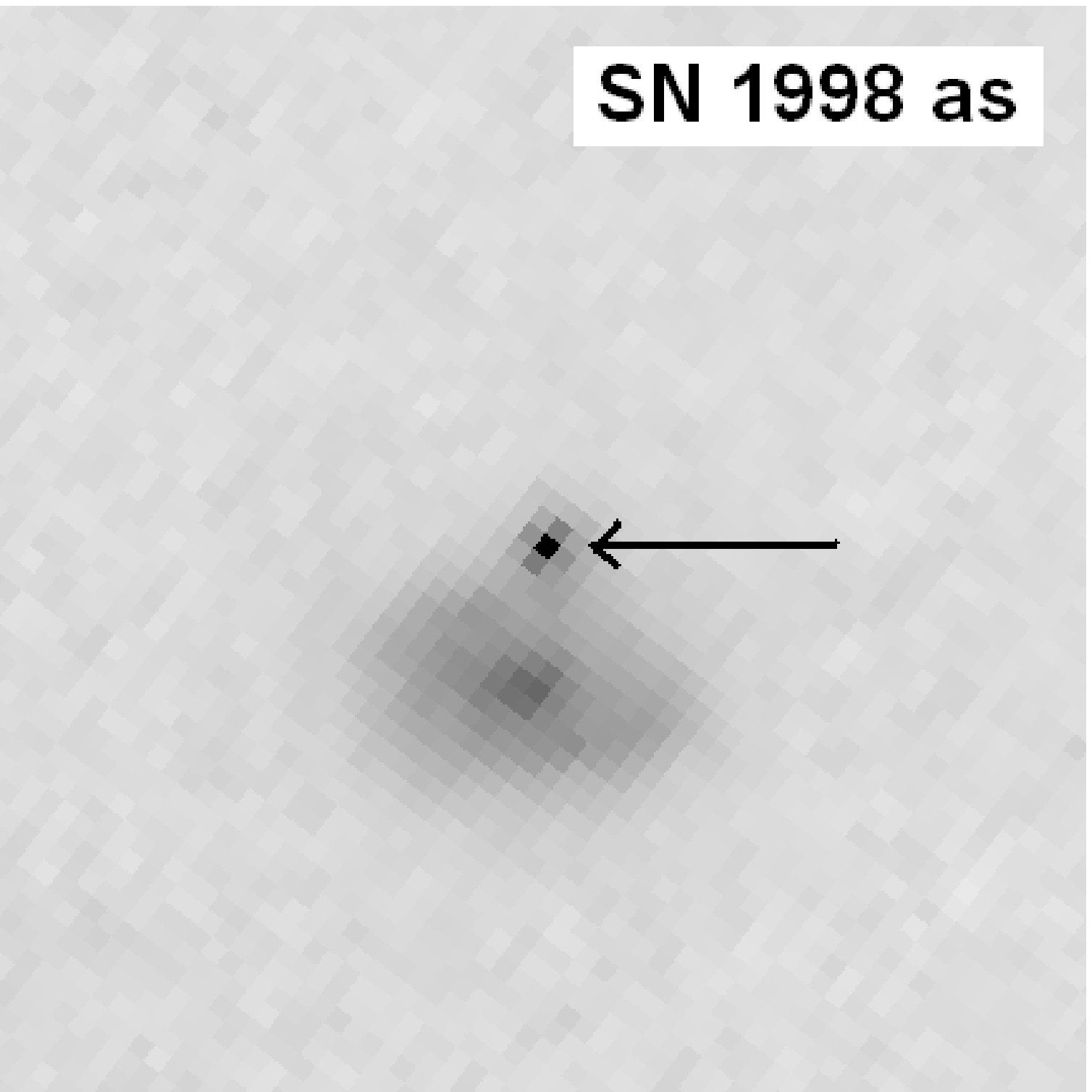}
 \includegraphics[width=7cm]{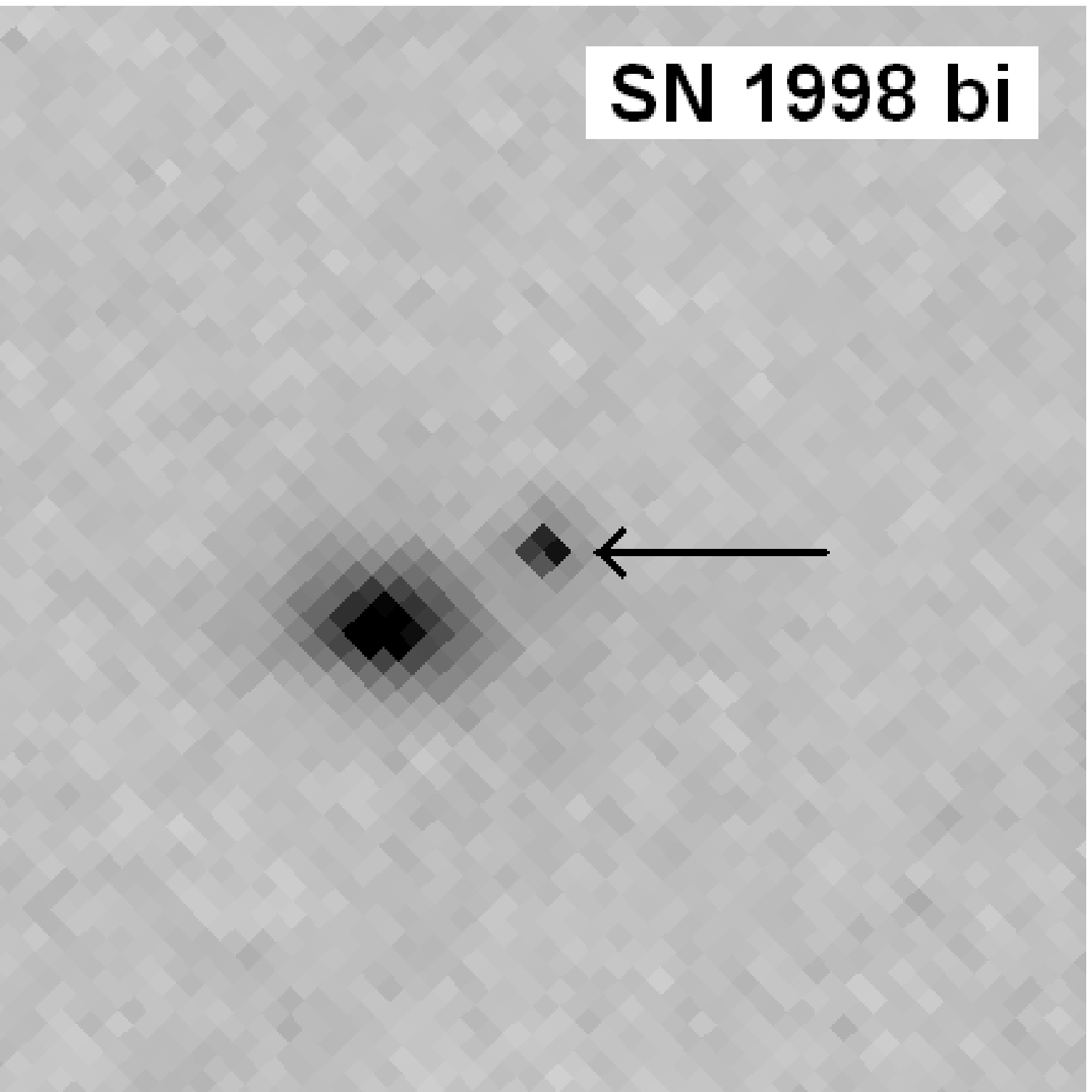}
 \includegraphics[width=7cm]{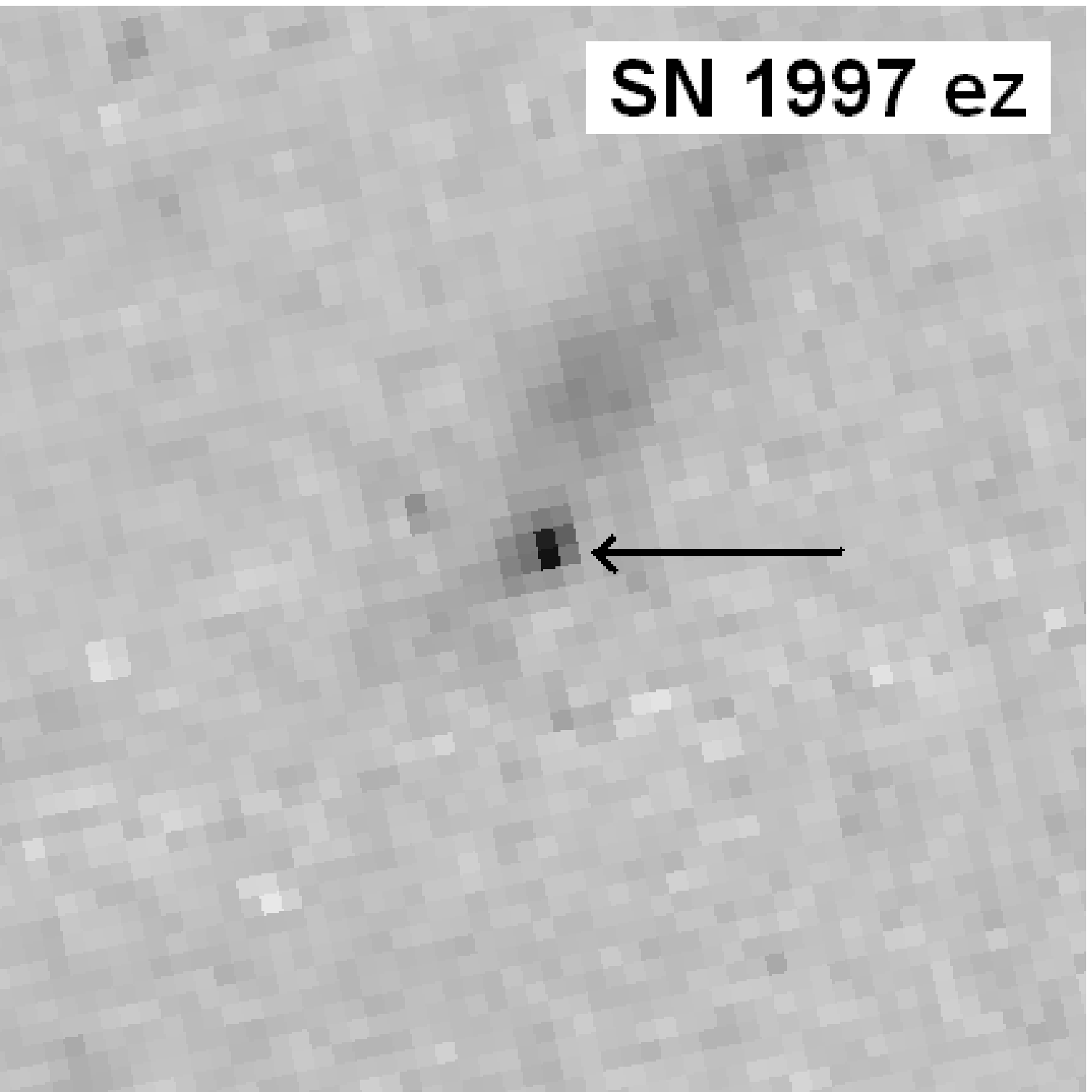}
 \includegraphics[width=7cm]{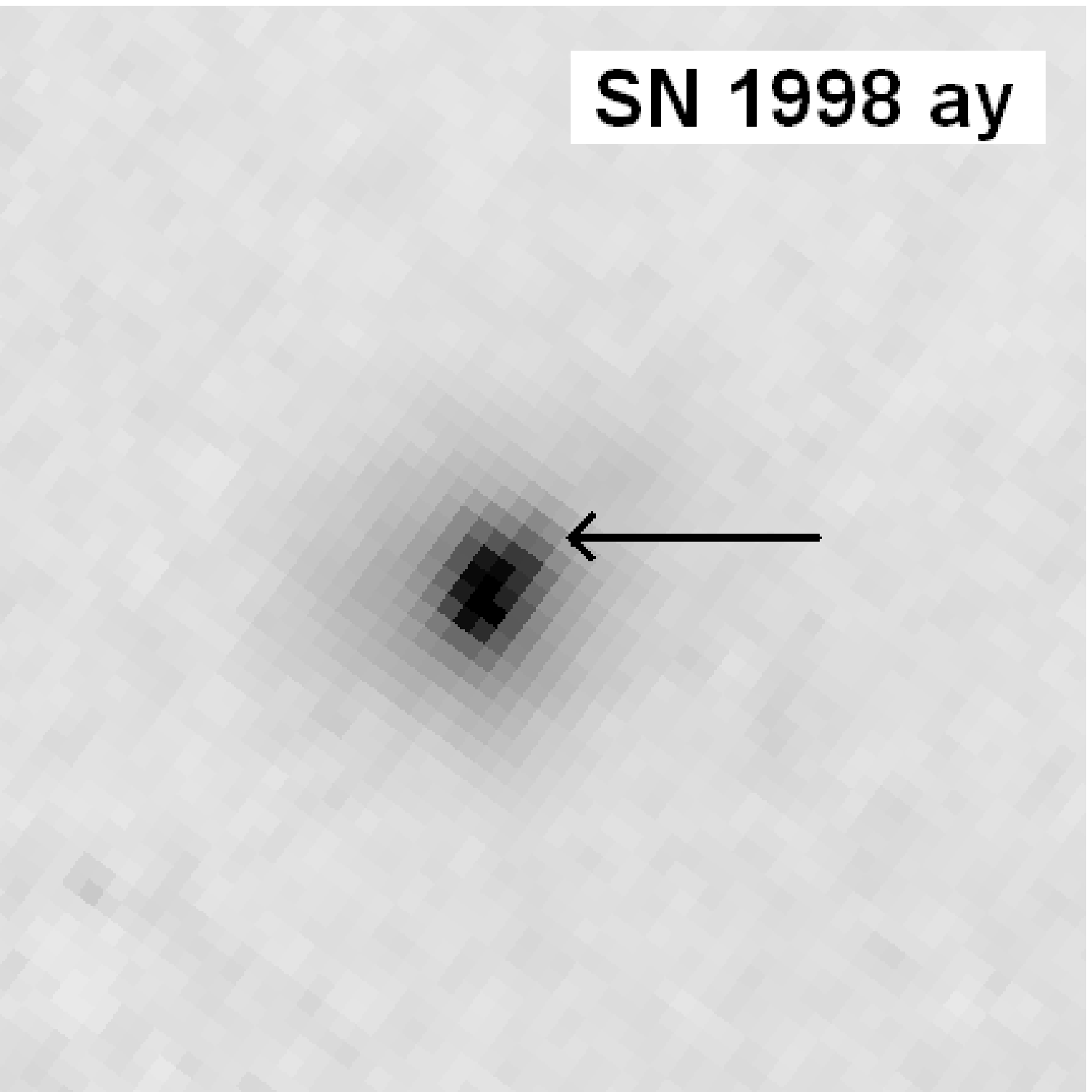}
 \caption{(From top to bottom, left to right) NICMOS images of
   SN~1997ek, SN~1998as, SN~1998bi, SN~1997ez, SN~1998ay and their
   respective host galaxies. SN~1998ay exploded close to the center of
   its host galaxy. An image of the host of SN~1998ay was taken about
   one year after the explosion, and it has been used for correcting
   the host galaxy contamination (see text for details). Each image is
   4$\times$4 arcsecs. North is up and east is on the left of each
   image.}
\label{nicmos}
\end{figure}

A summary of the observations is given in Table~\ref{table:phot}, including
the time since $B$-band maximum, exposure time in
seconds, magnitudes in the Vega system, and total uncertainties
(including both statistical and
systematic). Figure~\ref{nicmos} shows the infrared image 
taken at the brightest epoch observed. Each image, covering 4 $\times$ 4 
arcsecs, is built by co-adding several observations
for the purpose of displaying the position of the SN with respect
to the host galaxy. The actual photometry was performed on the
individual images.   
Table~\ref{table:SNe} reports redshifts, $B$-band stretch factors,
Milky Way reddenings, and color excesses for these SNe. 

\begin{deluxetable}{ccrccc}
\tablecolumns{6} 
\tablewidth{0pc}
\tabletypesize{\footnotesize}
\tablecaption{Summary of Infrared Images}
\tablehead{SN & MJD  & Epoch\tablenotemark{a} & Exp\tablenotemark{b} &
  Counts\tablenotemark{c} & Vega magnitudes\tablenotemark{d}}
\startdata 
SN~1997ek  & 50818.50 & -0.07 & 1026 & 0.483 $\pm$ 0.055 $\pm$ 0.014 & 23.058 $\pm$ 0.126 \\
           & 50824.34 & 5.77  & 2053 & 0.508 $\pm$ 0.020 $\pm$ 0.017 & 23.003 $\pm$ 0.055 \\
           & 50846.73 & 28.16 & 1280 & 0.441 $\pm$ 0.033 $\pm$ 0.049 & 23.157 $\pm$ 0.144 \\
           & 50858.69 & 40.12 & 2560 & 0.253 $\pm$ 0.017 $\pm$ 0.017 & 23.760 $\pm$ 0.101 \\
           & 50872.10 & 53.53 & 2560 & 0.171 $\pm$ 0.021 $\pm$ 0.030 & 24.186 $\pm$ 0.232 \\
SN~1997ez  & 50818.64 & 5.34  & 2052 & 0.466 $\pm$ 0.018 $\pm$ 0.011 & 23.097 $\pm$ 0.050 \\
           & 50824.55 & 11.25 & 2052 & 0.523 $\pm$ 0.019 $\pm$ 0.011 & 22.972 $\pm$ 0.047 \\
           & 50858.63 & 45.33 & 2560 & 0.231 $\pm$ 0.017 $\pm$ 0.007 & 23.859 $\pm$ 0.087 \\
           & 50872.27 & 58.97 & 2560 & 0.232 $\pm$ 0.018 $\pm$ 0.010 & 23.854 $\pm$ 0.096 \\
SN~1998bi  & 50922.51 & 10.70 & 3593 & 0.607 $\pm$ 0.014 $\pm$ 0.015 & 22.810 $\pm$ 0.037 \\
SN~1998as  & 50909.35 & 13.23 & 5134 & 0.882 $\pm$ 0.013 $\pm$ 0.026 & 22.404 $\pm$ 0.036 \\
           & 50921.27 & 25.15 & 5134 & 0.879 $\pm$ 0.013 $\pm$ 0.026 & 22.408 $\pm$ 0.037 \\
SN~1998ay  & 50910.33 & 14.15 & 4105 & 0.554 $\pm$ 0.017 $\pm$ 0.012 & 22.909 $\pm$ 0.040 \\
           & 50920.28 & 24.10 & 4105 & 0.409 $\pm$ 0.015 $\pm$ 0.009 & 23.239 $\pm$ 0.047 \\
\enddata 
\tablenotetext{a}{Days since B-band maximum in the observer frame.}
\tablenotetext{b}{Exposure time in seconds.}
\tablenotetext{c}{Counts (ADUs) per second. For conversion to electrons per second one can
use the gain of 5.4 e-/ADU \cite{BarkerDahlen}. The first uncertainty
  shown is statistical and the second one is systematic.}
\tablenotetext{d}{Infrared Photometry. The uncertainty includes both
  statistical and systematic components, as well as a 2\% zeropoint
  calibration uncertainty.} 
\label{table:phot}
\end{deluxetable}

\begin{deluxetable}{ccccr}
\tablecolumns{5} 
\tablewidth{0pc}
\tablecaption{Supernova Name, Redshift, $B$-band Stretch Factor,
Milky Way Reddening, and Color Excess.}
\tablehead{SN & $z$ & $s_B$ & $E(B-V)_{MW}\tablenotemark{a}$ & $E(B-V)$}
\startdata
SN~1997ek\tablenotemark{b} & 0.862 & 1.059 $\pm$ 0.071 & 0.042 & $-0.04 \pm 0.05$\\
SN~1997ez  &  0.776  &  1.094 $\pm$ 0.038 & 0.026 & $0.20 \pm 0.08$\\
SN~1998as  &  0.356  &  0.931 $\pm$ 0.018 & 0.037 & $0.19  \pm 0.04$\\
SN~1998ay  &  0.638  &  1.050 $\pm$ 0.045 & 0.035 & $0.09 \pm 0.09$\\
SN~1998bi  &  0.75   &  0.975 $\pm$ 0.031 & 0.026 & $0.16 \pm 0.04$\\
\enddata 
\tablenotetext{a}{Milky Way reddening given by \citet{1998ApJ...500..525S}}
\tablenotetext{b}{The color excess for SN~1997ek is estimated excluding the data
  point at day $\sim 20$. See text for details.}
\label{table:SNe}
\end{deluxetable}

\begin{figure}
  \includegraphics[width=7.5cm]{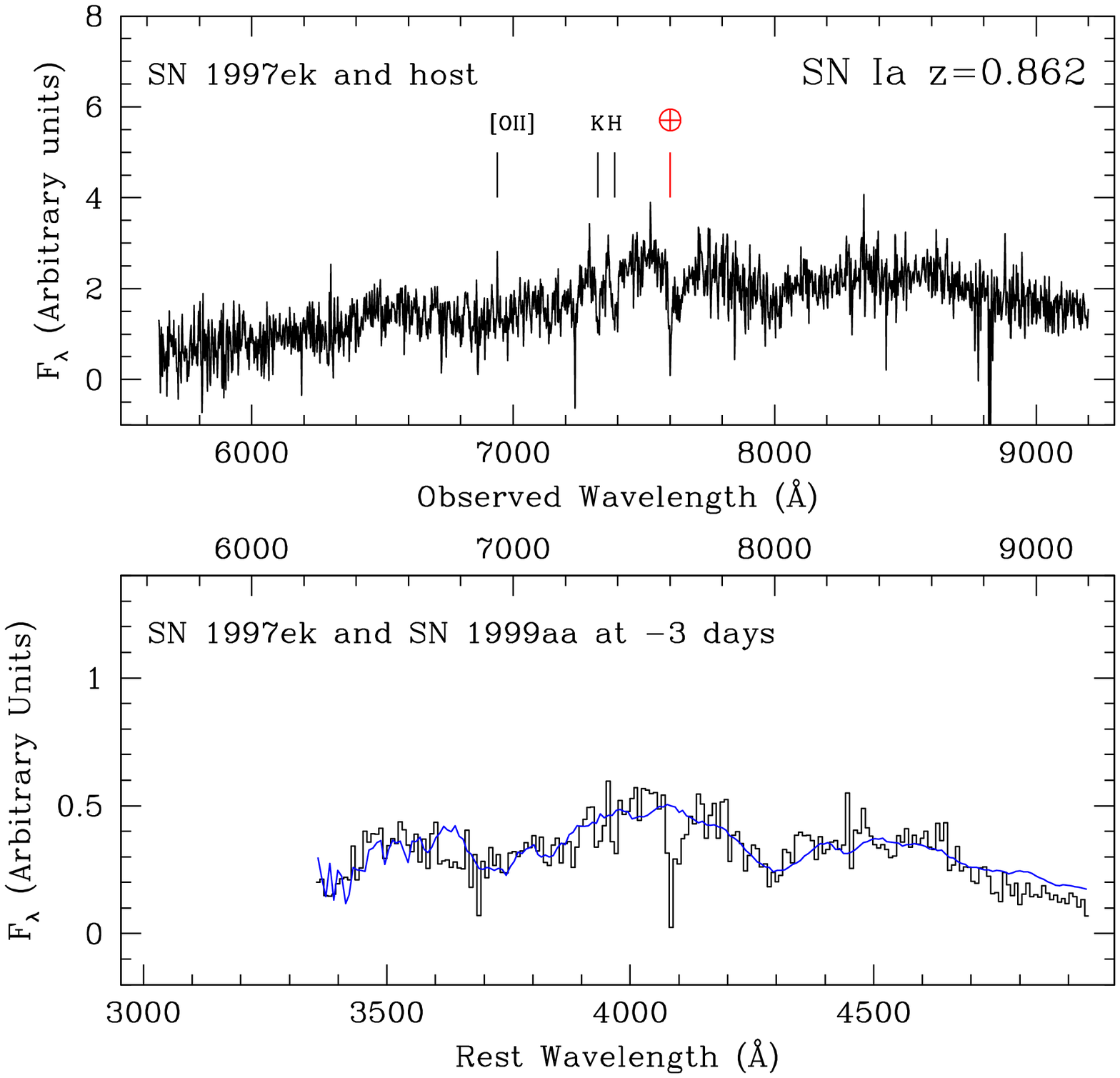}
  \includegraphics[width=7.5cm]{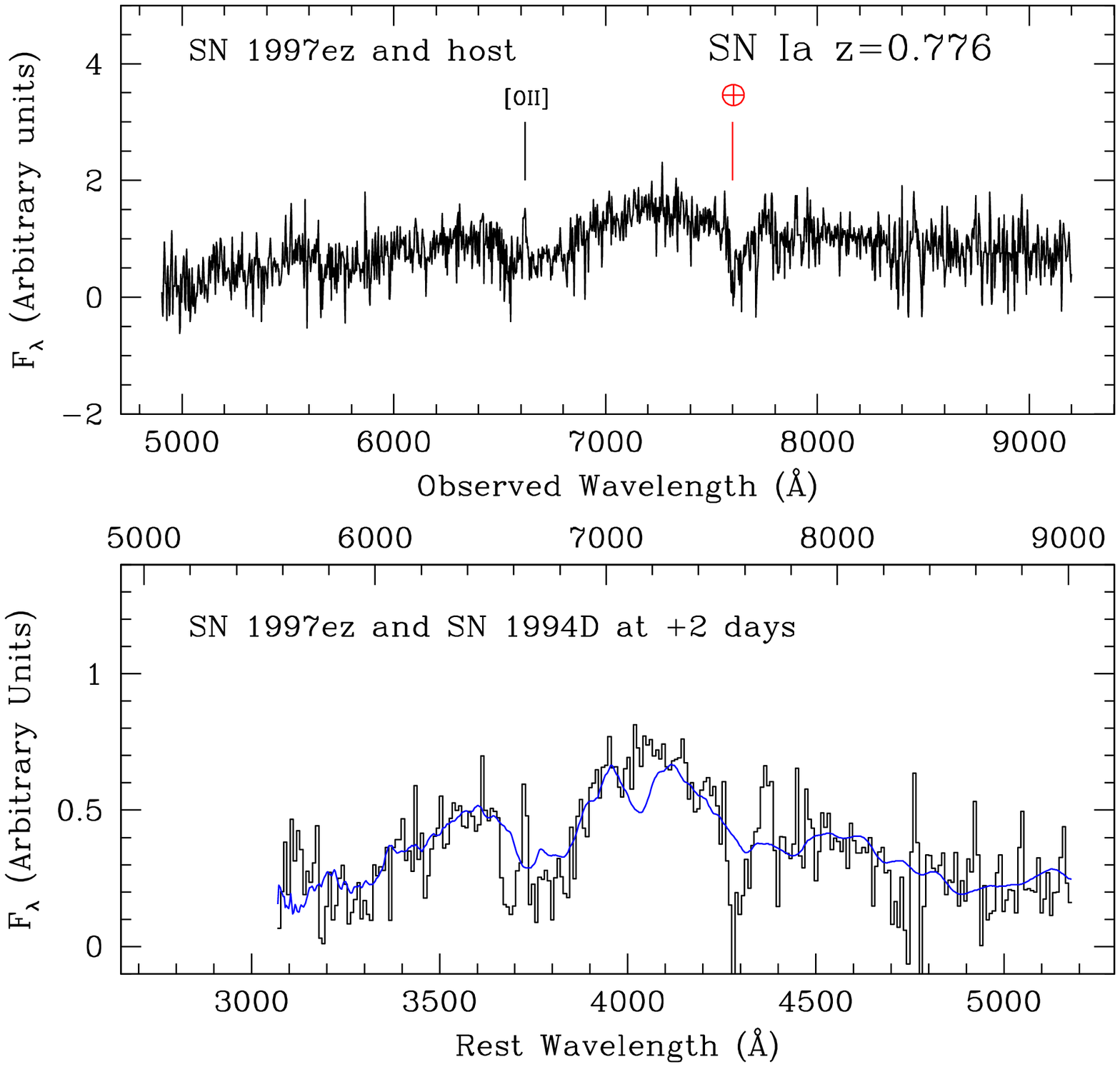}
  \includegraphics[width=7.5cm]{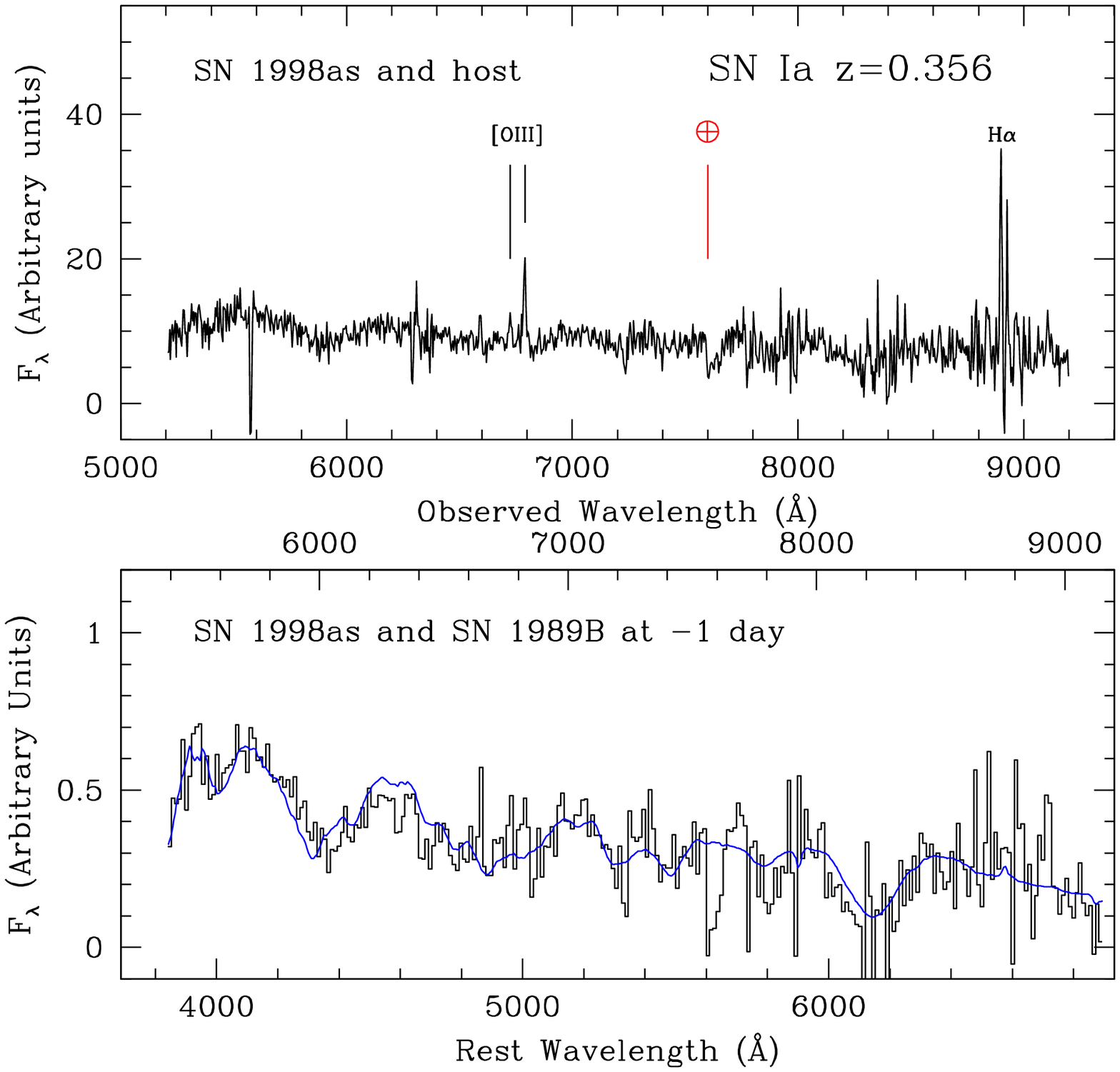}
  \includegraphics[width=7.5cm]{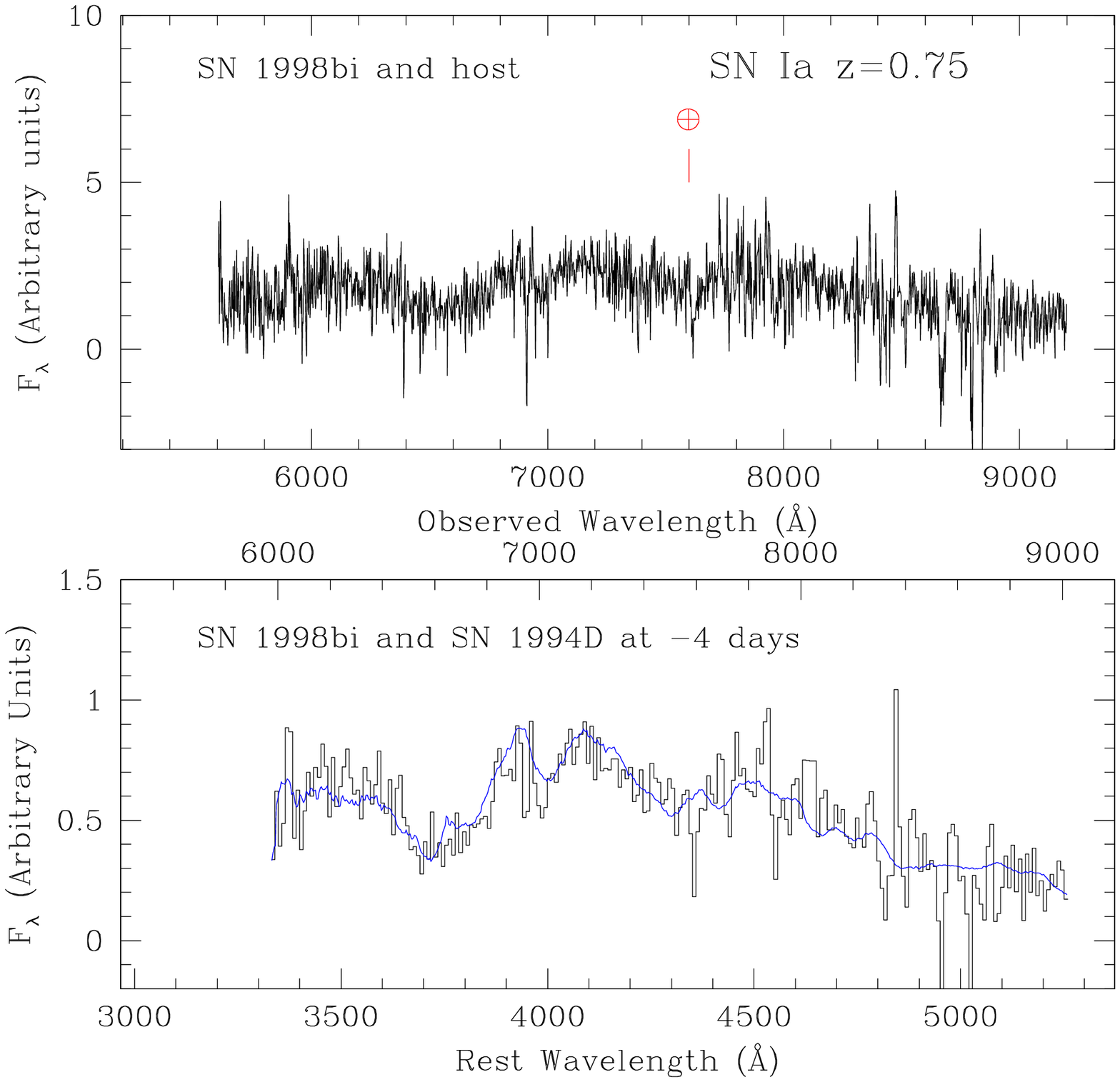}
  \center\includegraphics[width=7.5cm]{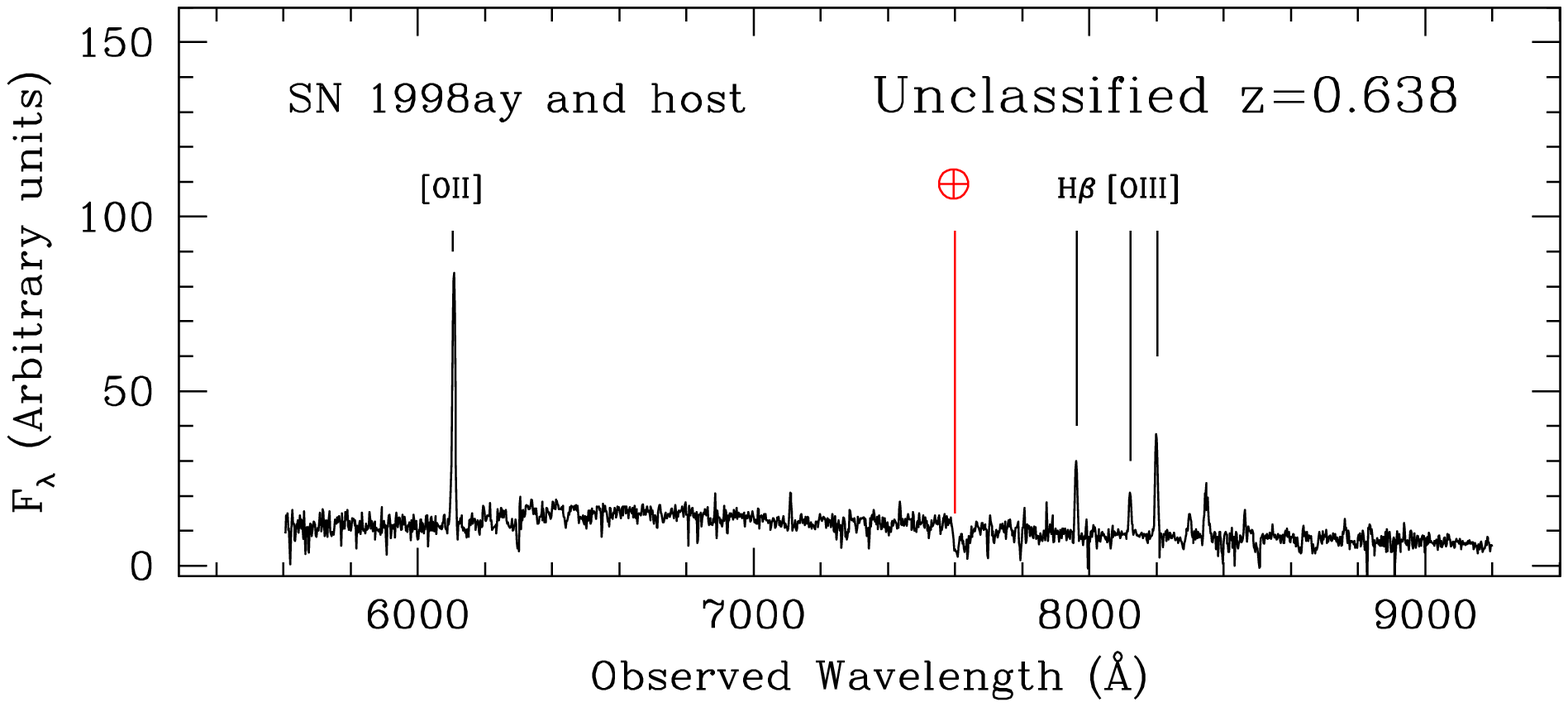}
\caption{\small Confirmation spectroscopy. With the exception of SN~1998ay, each SN is 
shown twice. The observed spectrum, including light from the host galaxy 
is plotted in the upper panels. In these spectra, the most prominent host 
galaxy lines, if any, are marked. The red circled crosses mark regions of 
strong telluric absorption. In the lower panels, the binned host galaxy 
subtracted spectra (the black histogram) is shown together with the best 
fitting local SN~Ia (the blue continuous line). The spectrum 
of SN~1998ay is plotted only once as light from the host galaxy dominates 
the spectrum.}
\label{spectra}
\end{figure}

\section{Spectroscopic confirmation}
\label{sec:spectra}

Optical spectra of SN~1997ek, SN~1997ez, SN~1998ay, and SN~1998bi 
were obtained at Mauna Kea (Keck II telescope) in December 1997 and
April 1998. A spectrum of SN~1998as was obtained at the European
Southern Observatory 3.6-m telescope in March 1998. All supernovae were
within 2.5 arcsecs of the host galaxy center. Since
\citet{2003ApJ...598..102K}, the spectra have been re-analysed and the
redshifts re-determined. In general, redshifts are measured from host galaxy
emission and absorption lines. The one exception is SN~1998bi,
for which the supernova was used. Within the uncertainties, the
redshifts agree with those in \citet{2003ApJ...598..102K}. 

The extracted spectra, which consist of light from both the supernova
and the host galaxy, were fitted with the algorithm described in
\citet{2005ApJ...634.1190H}. The type, the confidence that the
supernova is a SN~Ia, and the mean rest-frame spectral age from the
five best fits are shown in Table ~\ref{table:spectra}. The extracted
spectra, the host-subtracted spectra, and the best
matching local SN Ia are shown in Figure \ref{spectra}.
With the exception of SN~1998ay, all of the SNe could be identified as
either a certain SN~Ia (confidence index of 5) or a highly probable
one (confidence index of 4) . The spectrum of SN 1998ay is heavily
contaminated with host galaxy light so a secure classification based
on the spectrum is not possible. However, as we show in the next
section, the light curve and color of SN~1998ay are consistent with
those of a SN~Ia.  

\begin{deluxetable}{llllllllll}
\rotate
\tablecolumns{10} 
\tablewidth{0pc}
\tabletypesize{\footnotesize}
\tablecaption{Summary of Spectroscopic Observations.}
\tablehead{SN & Telescope & Instrument/Setup\tablenotemark{a} & Date(UT) & Exposure\tablenotemark{b} & Best Match & Host Lines & Epoch\tablenotemark{c} & Index & Classification}
\startdata 
SN~1997ek  &  Keck II  & LRIS/400/8500/1  & 02/01/1998  & 3x1500 & SN~1999aa -3 days & [\ion{O}{2}],H,K & -1.8 (4.1) & 4 & SN Ia\\
SN~1997ez  &  Keck II  & LRIS/300/5000/1  & 31/12/1997  & 3x1800 & SN~1994D +2 days  &  [\ion{O}{2}] & +0.4 (6.4) & 4 & SN Ia\\
SN~1998as  &  ESO 3.6m & EFOSCII/R300/1.5 & 28/03/1998  & 2x1800 & SN~1989B -1 day  &   H$\beta$, [\ion{O}{3}], H$\alpha$ & -1.8 (4.3) & 5 & SN Ia\\
SN~1998ay  &  Keck II  & LRIS/400/8500/1  & 03/04/1998  & 2x1200 & No match  & [\ion{O}{2}], H$\beta$, [\ion{O}{3}] &    & 2   & SN \\
SN~1998bi  &  Keck II  & LRIS/400/8500/1  & 01/04/1998  & 1x1500 & SN~1994D -4 days &   & -3.4 (3.9)  & 5  & SN Ia\\
\enddata
\tablenotetext{a}{For the LRIS
  observations, 400/8500/1 means that the observations were done with
  the 400/8500 grating and the 1\arcsec\ slit. For the EFOSC II observations,
  R300/1.5 means that the observations were done with the R300 grism and
  the 1\farcs5 slit.}
\tablenotetext{b}{Exposure time expressed in seconds}
\tablenotetext{c}{Average rest-frame epoch of the five best fits, with the
  rms value reported in parenthesis} 
\label{table:spectra}
\end{deluxetable}

\section{Supernova colors}
\label{sec:colour}

We use the infrared data together with the optical data presented by
 \citet{2003ApJ...598..102K} to compare 
photometric properties of distant SNe~Ia to those of nearby SNe~Ia. 
Intrinsic color properties of nearby SNe~Ia have been extensively studied by 
\citet{2008A&A...487...19N}.
The reported color evolution with stretch
and supernova epoch are used in the analysis presented here,
both as a model for the intrinsic colors and for the 
$K$-corrections \citep{1996PASP..108..190K}. 
\citet{2002PASP..114..803N} showed that
the latter depends mainly on the supernova color rather than on
changes in individual spectral features. Thus we use the 
spectral templates developed by \citet{2007ApJ...663.1187H} modified
by the color-stretch relation determined in
 \citet{2008A&A...487...19N} before computing the $K$-corrections.  
To compare the observed colors of the SNe in this data-set with 
color templates derived for low-z supernovae we need to know three
parameters: the epochs of our observations with respect to restframe 
$B$-band maximum, the stretch $s$ of the rest-frame $B$-band lightcurve 
and the color excess, $E(B-V)$. To obtain these parameters, the lightcurves
need to be fitted iteratively, since the $K$-corrections needed for the 
fits depend on both $s$ and $E(B-V)$.
\noindent We use the following procedure:

\begin{enumerate}

\item Compute $K$-corrections using the $s=1$ spectral template
  corrected for the color excess, i.e. warped by the extinction law by
  \citet{1989ApJ...345..245C} (CCM) with the values of $E(B-V)$ for
  each SN from \citet{2003ApJ...598..102K} using a total to
  selective extinction ratio $R_V=1.75$, which has been found to give
  a good empirical description of reddening in low-z SNIa samples
  \citep[][and references therein]{2008A&A...487...19N}.
 
 \item Fit the optical light curves to determine the time of $B$-band
  maximum and the stretch factor, $s$, for each SN. 

\item Compute the intrinsic colors corresponding to
  $s$ following the color-stretch relation described in
  \citet{2008A&A...487...19N} 

\item Warp the spectral templates\footnote{We use a cubic spline
    interpolation of the ratio between the synthetic photometry and
    the new photometry at the effective wavelengths. The $B$-band
    light curve by \citet{2001ApJ...558..359G} is assumed in order to
    compute all the $UVRI$ bands given the colors $U-B$, $B-V$, $V-R$
    and $R-I$ (see \citet{2008A&A...487...19N} for details).} to match
  the expected colors for the fitted $s$. We also redden the spectral
  template assuming as a first approximation the color excess $E(B-V)$
  determined by \citet{2003ApJ...598..102K}. Thus, we use the modified
  spectral template to compute $K$-corrections, while keeping the
  measured colors unchanged.

\item Fit the optical light curves again using updated K-corrections

\item Build the color models  using the fitted $I$-band light curve model
  interpolated to the epochs of the $R$-band observations
  to compute $R-I$, and to the epochs of the $F110W$-band 
  observations to compute $I-F110W$.
\end{enumerate}

\noindent Note that, in the previous steps we have not corrected the
observed photometry for the extinction in the Milky Way or in the host
galaxy. Instead, we have modified the spectral template needed for
fitting the data.  We use a grid of values of $E(B-V)$ in steps of
0.01 mag with a spectral template reddened according to the CCM
extinction law.  We assume a value of the parameter $R_V \equiv
A_V/E(B-V) = 1.75$ as in \citet{2008A&A...487...19N}. We note that a
recent work by \citet{2008ApJ...686L.103G} provides a potential
explanation for the empirically found low average value of the total
to selective extinction ratio in Type Ia supernovae as being a result of
interaction with dust in the circumstellar environment, (see also
\citet{2005ApJ...635L..33W}).  Step 4 above is repeated until the fit
parameters do not change outside the estimated uncertainties.



Finally, we compute synthetic photometry and build model color curves in the
observer-frame filters, $R-I$ and $I-F110W$, for each SN,
reddened by the Milky Way extinction
\citep{1998ApJ...500..525S}, using the standard value $R_V = 3.1$. 
The estimated color excess, $E(B-V)$, giving the best fit
to the observed optical data, $R-I$, is reported in
Table~\ref{table:SNe}.   
These values are then used to redden the color curves $I-F110W$ before
comparing with the observed infrared data.

\begin{figure}
  \includegraphics[width=8cm]{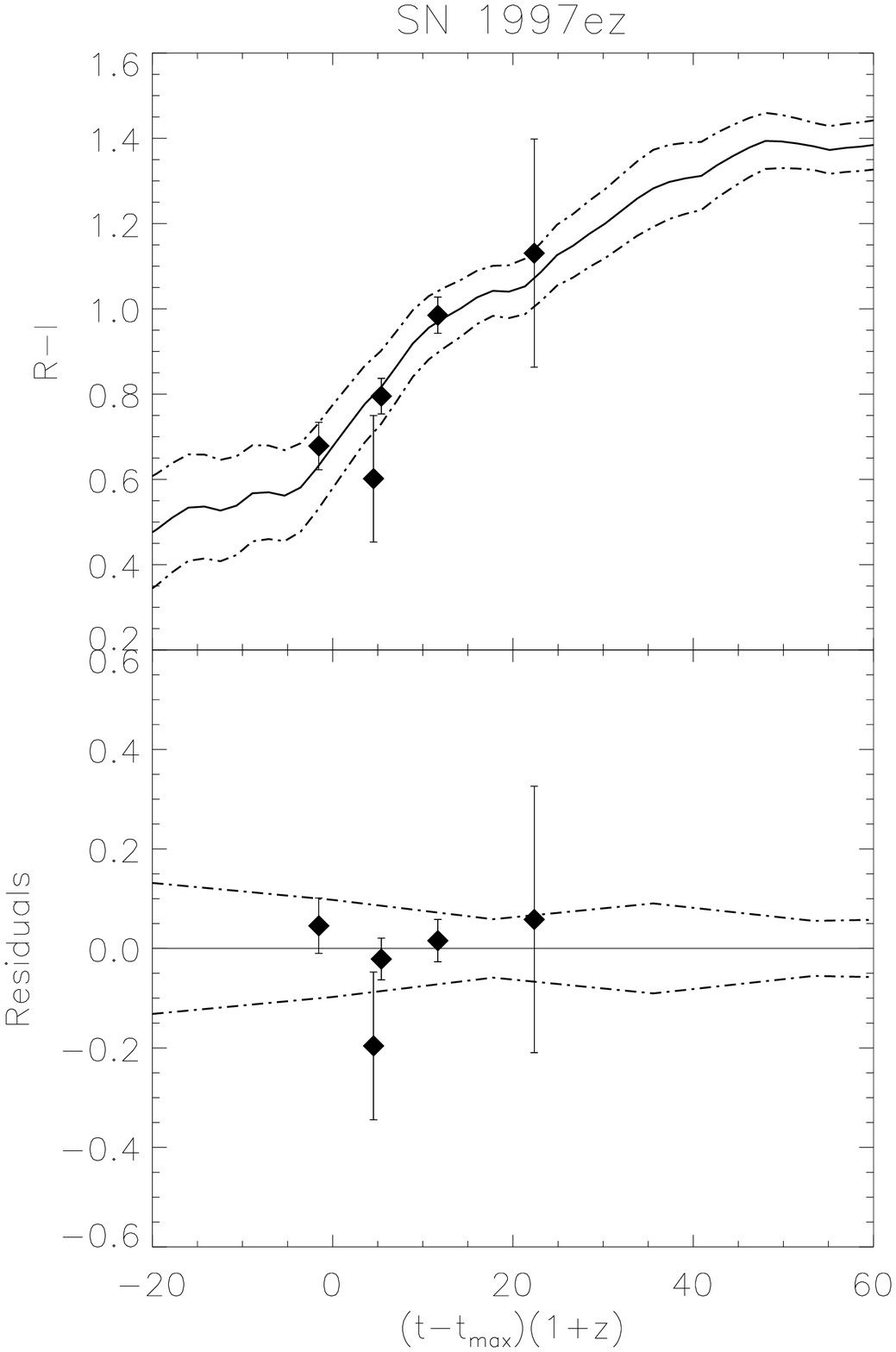}
  \includegraphics[width=8cm]{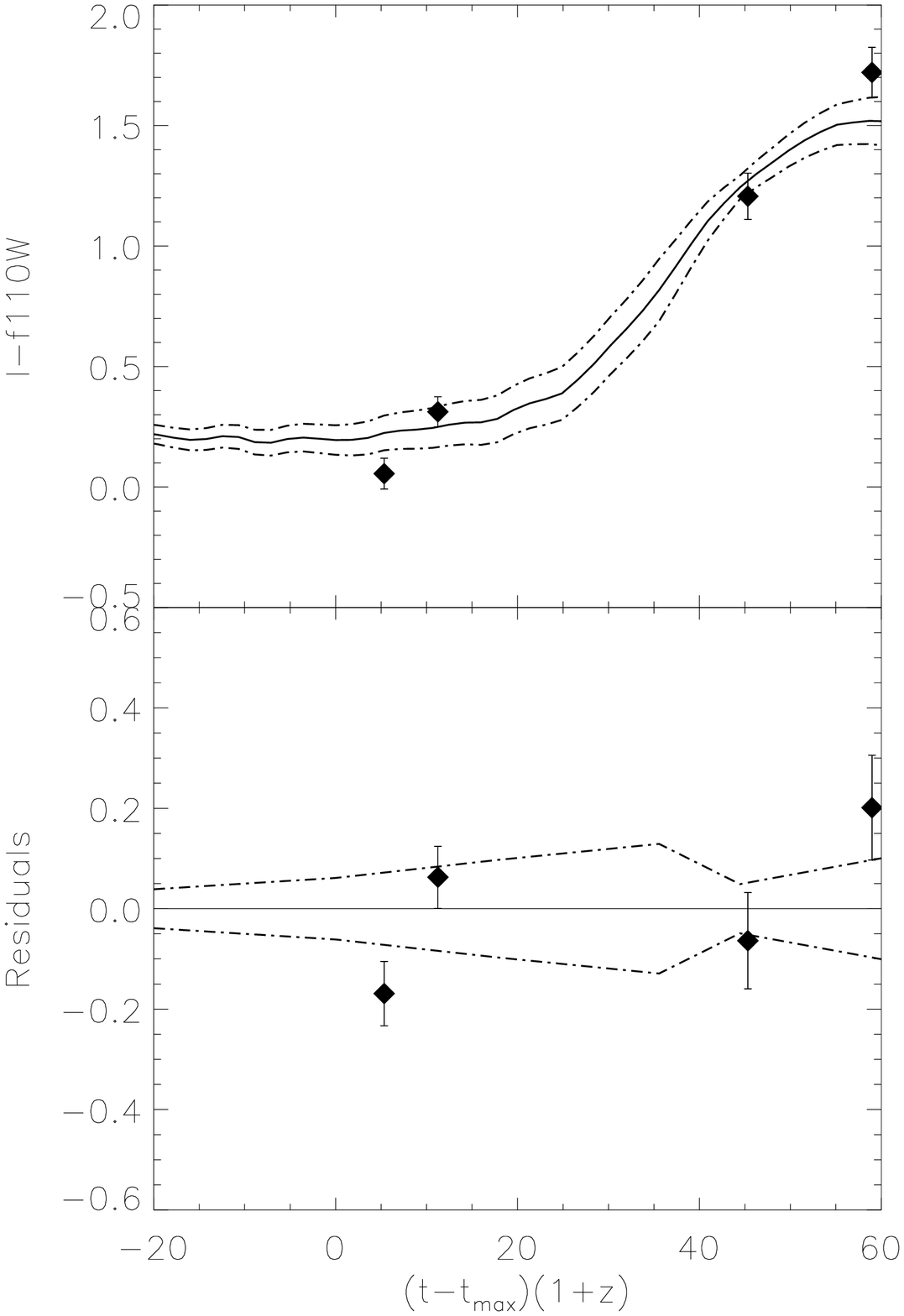}
  \includegraphics[width=8cm]{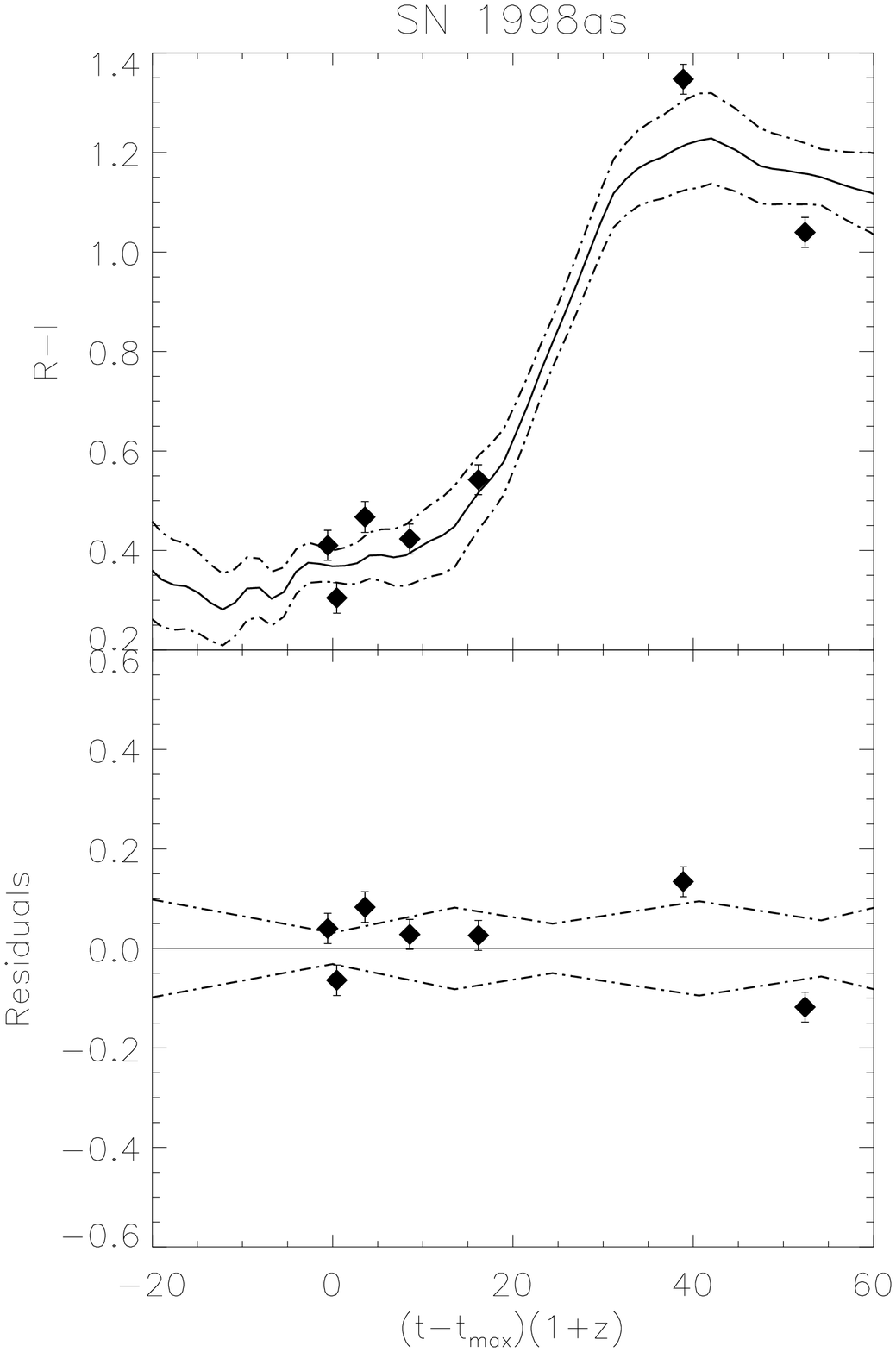}  
  \includegraphics[width=8cm]{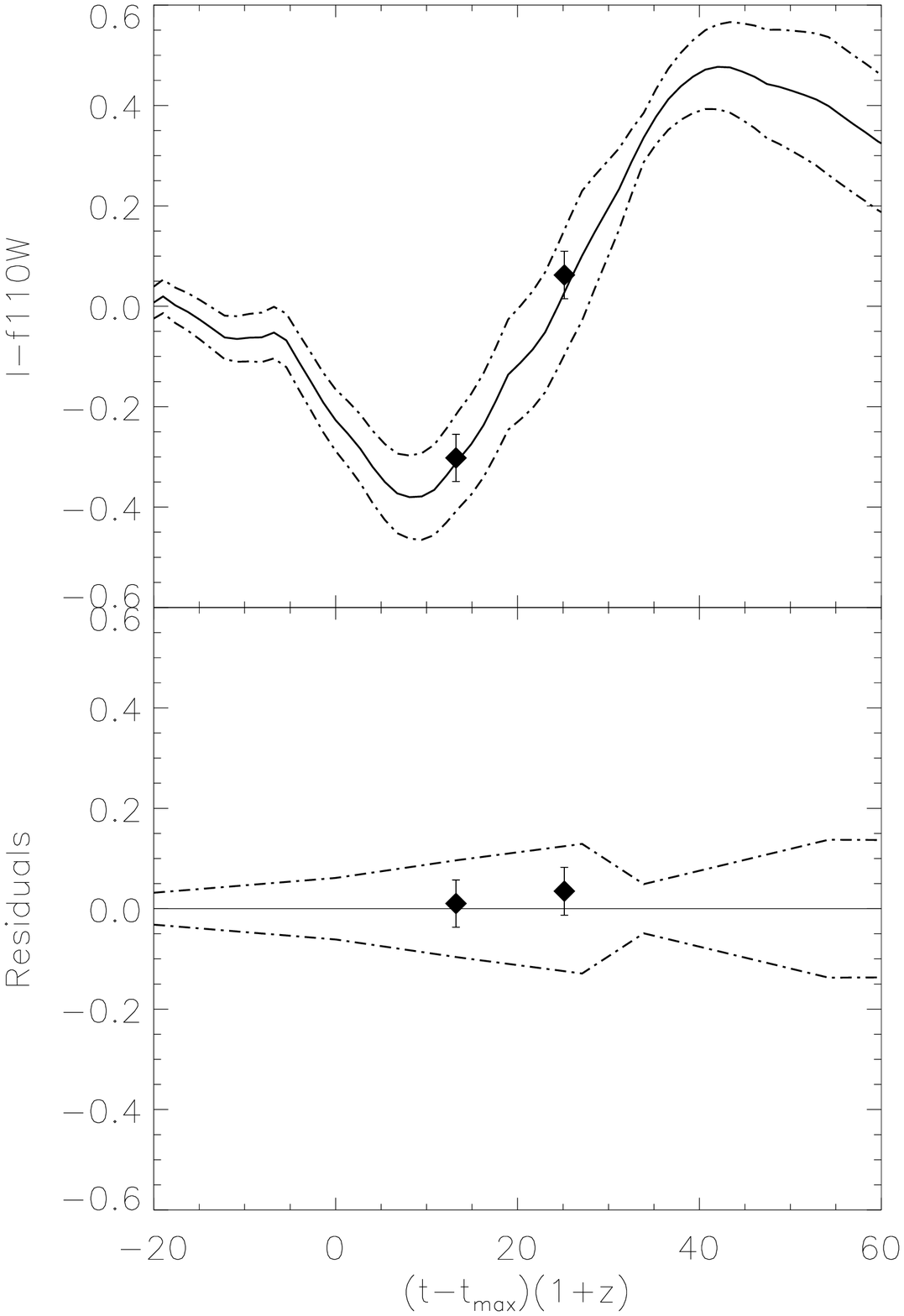}
  \caption{Observer-frame color for each SN. The color curves are
  obtained by synthetic photometry of the spectral templates (see text
  for details). The dashed lines show the intrinsic dispersion around
  the color curves measured on nearby SNe~Ia
  \citep{2008A&A...487...19N}.}
  \label{colors}
\end{figure}

\begin{figure}
  \includegraphics[width=8cm]{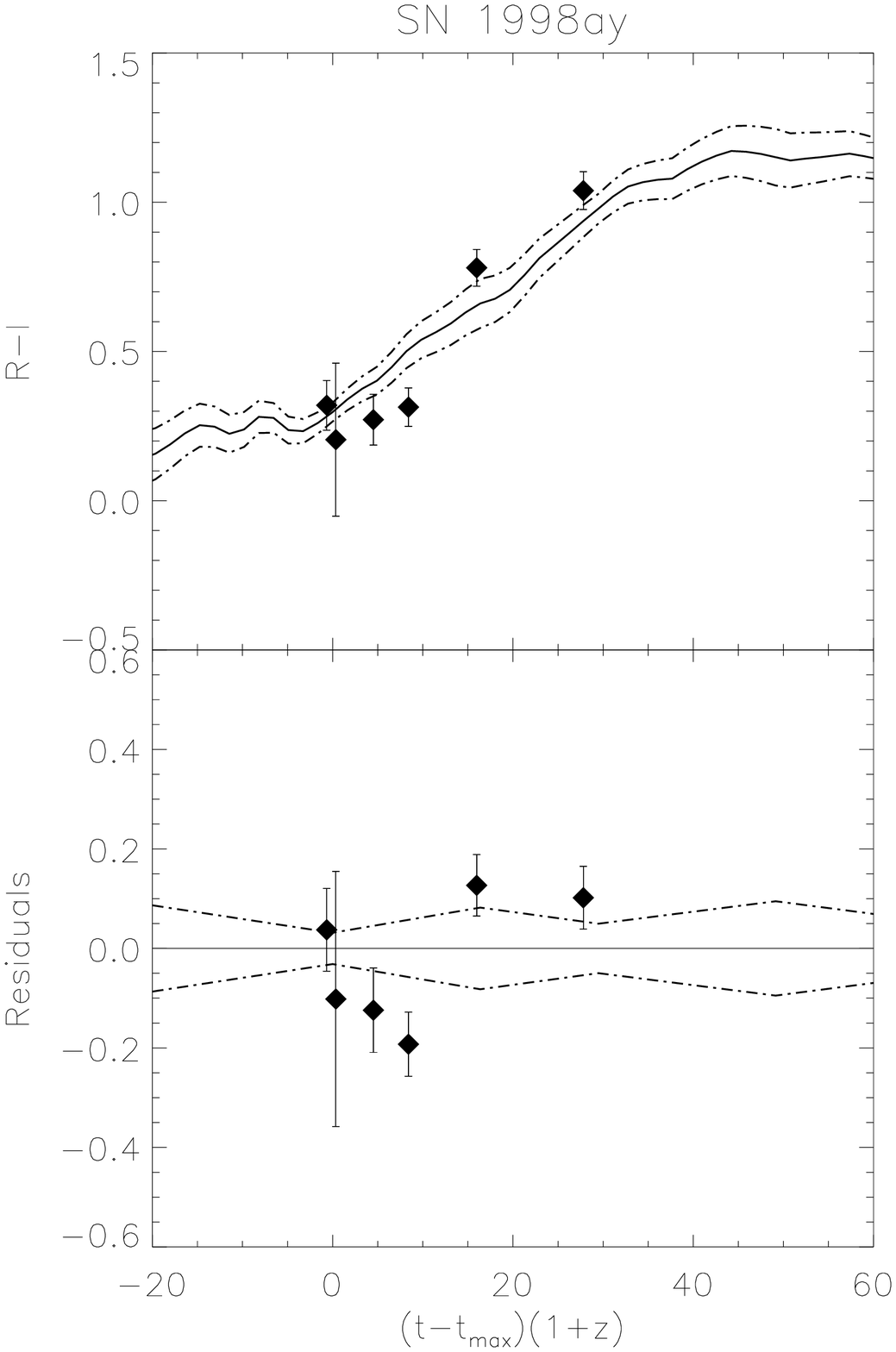}
  \includegraphics[width=8cm]{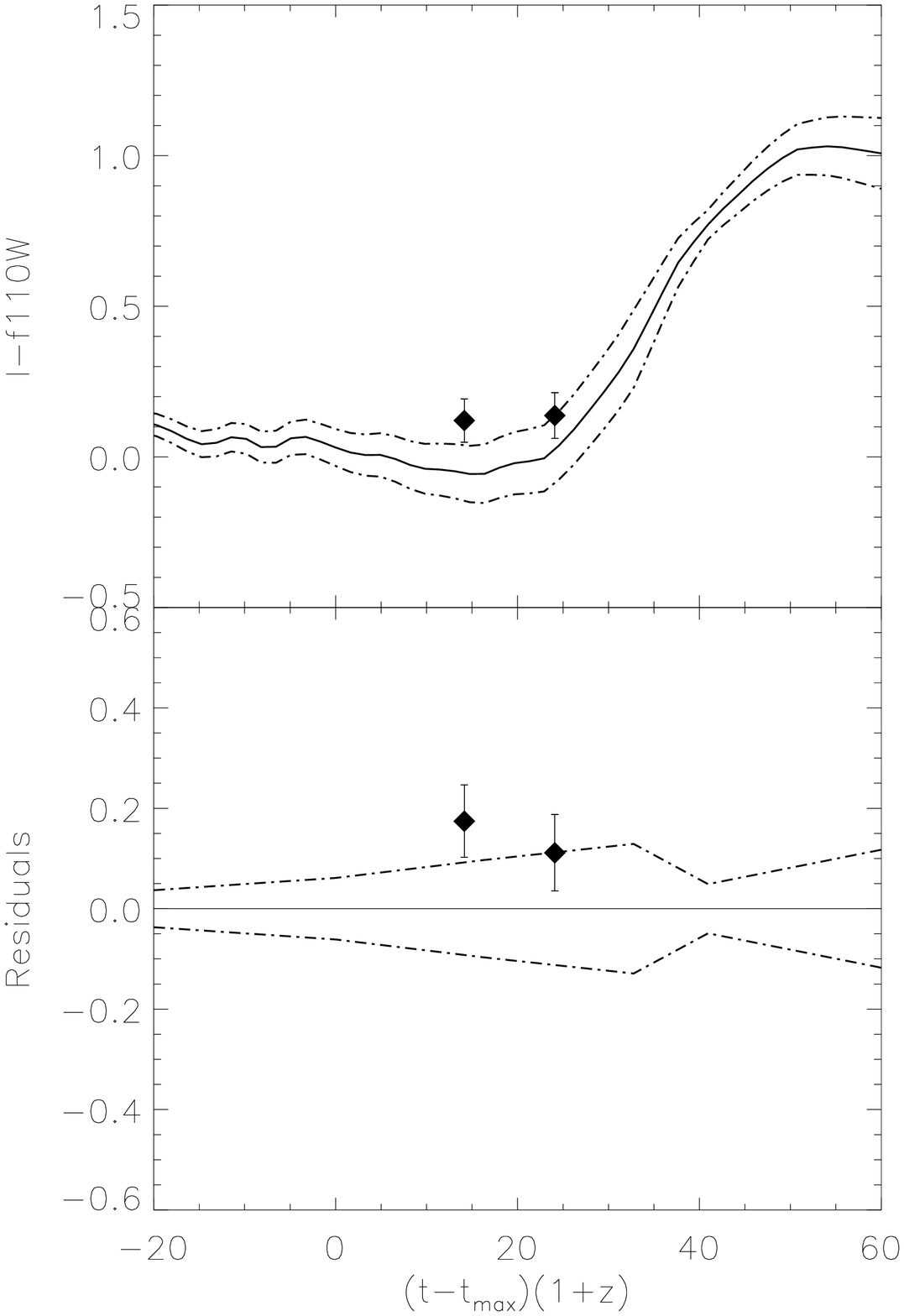}
  \includegraphics[width=8cm]{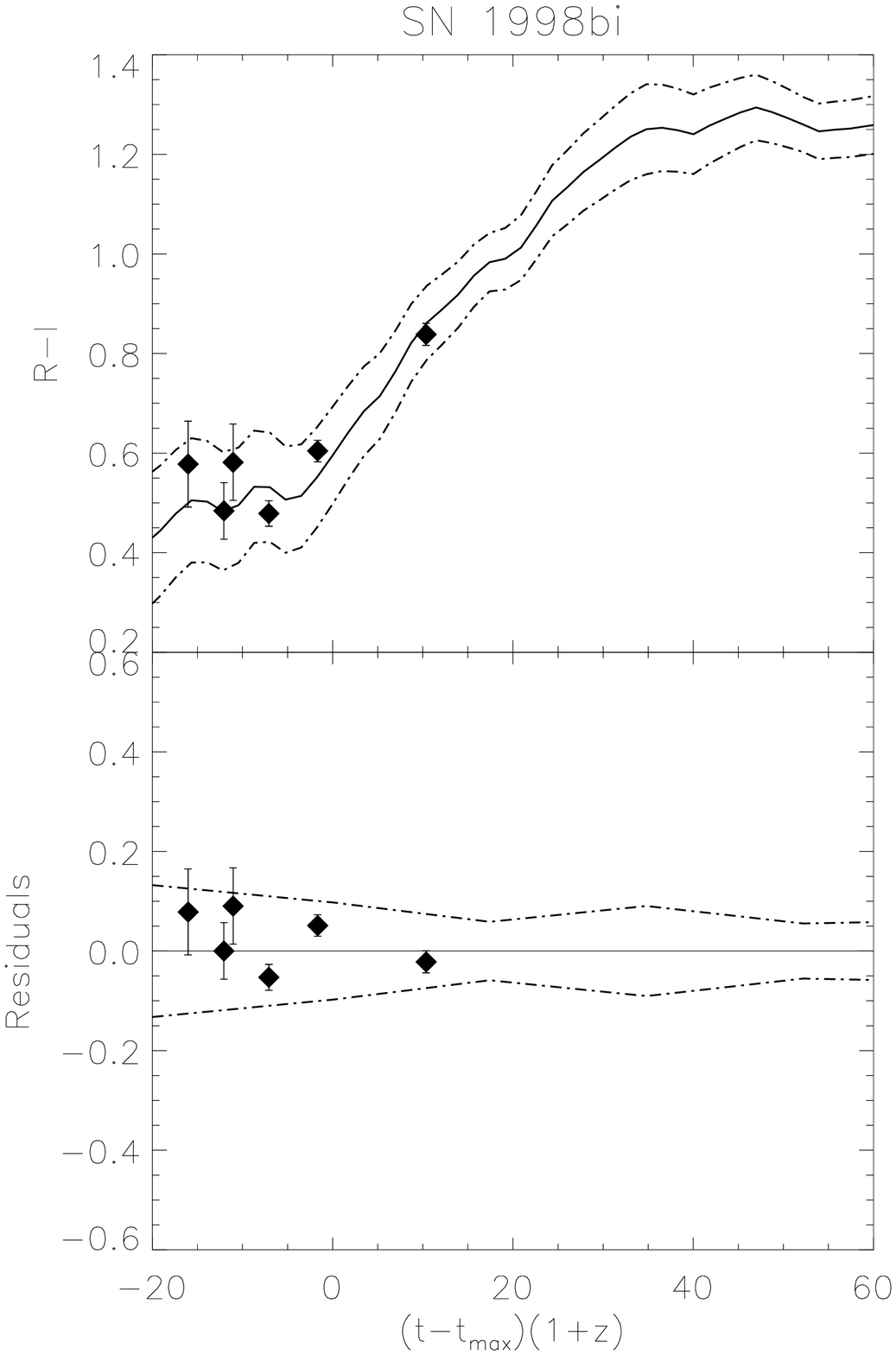} 
  \includegraphics[width=8cm]{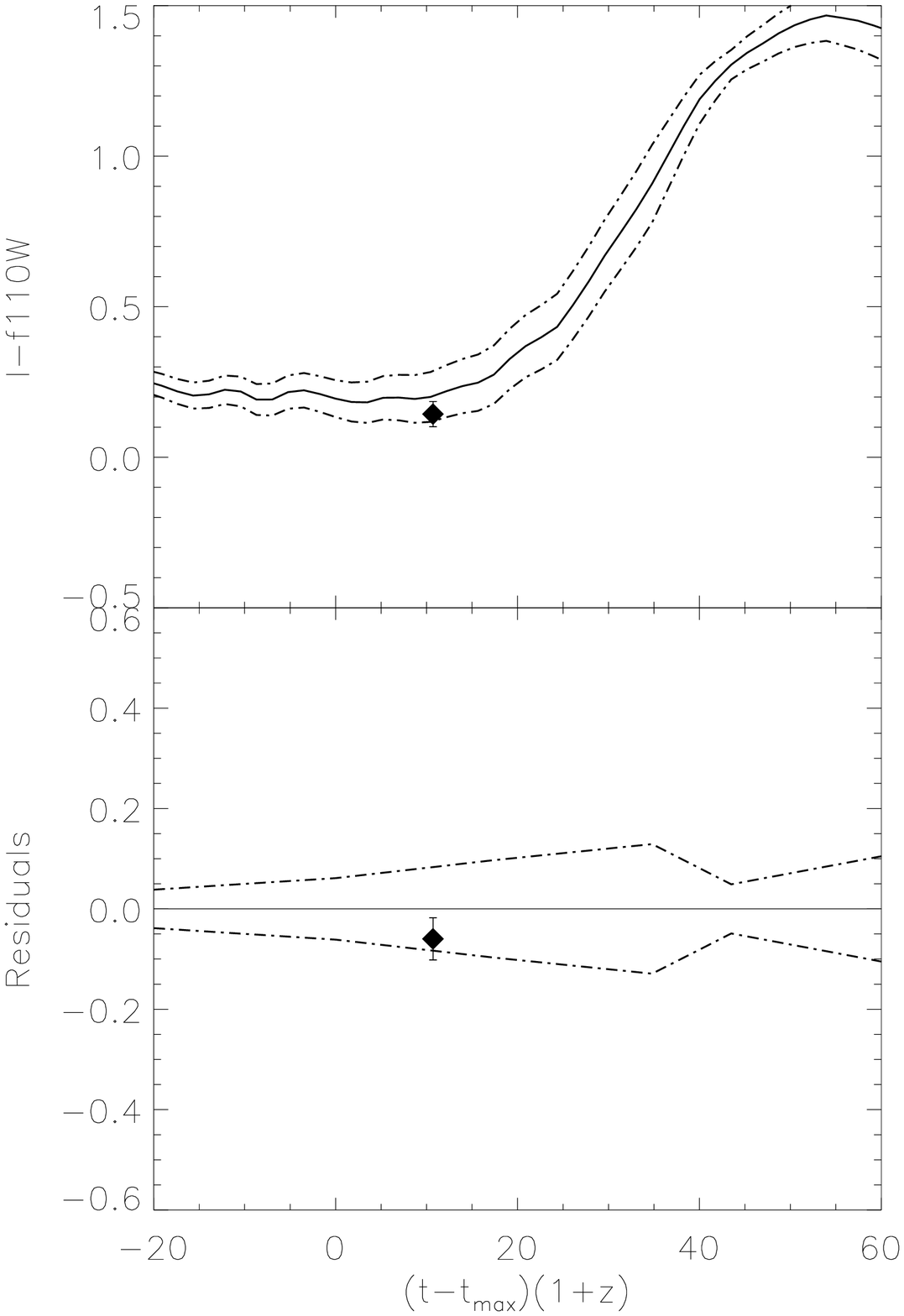}
  \caption{Observer-frame color for each SN. The color curves are
  obtained by synthetic photometry of the spectral templates (see text
  for details). The dashed lines show the intrinsic dispersion around
  the color curves measured on nearby SNe~Ia
  \citep{2008A&A...487...19N}.}
  \label{colors1}
\end{figure}

\begin{figure}
  \includegraphics[width=8cm]{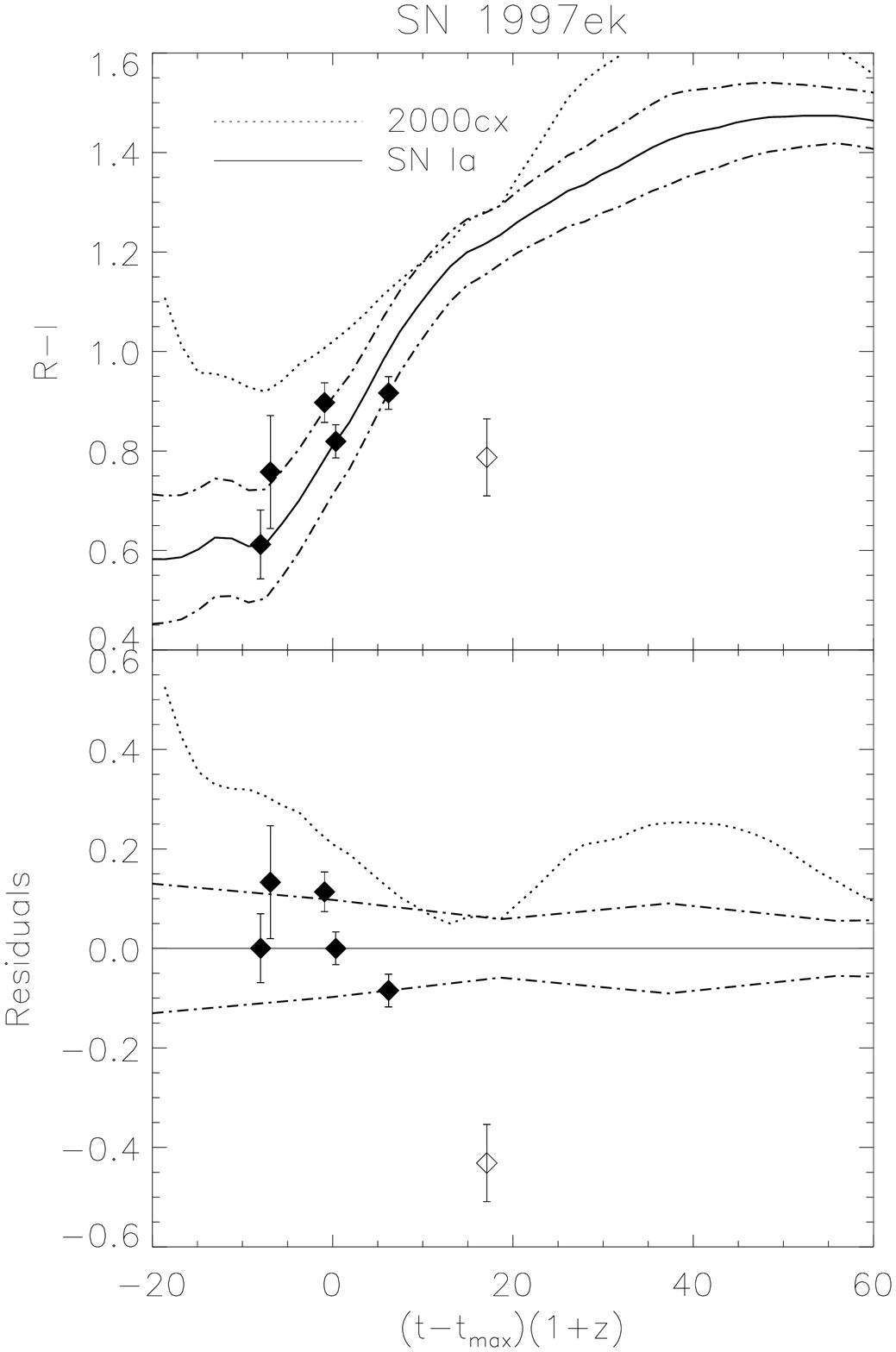}
  \includegraphics[width=8cm]{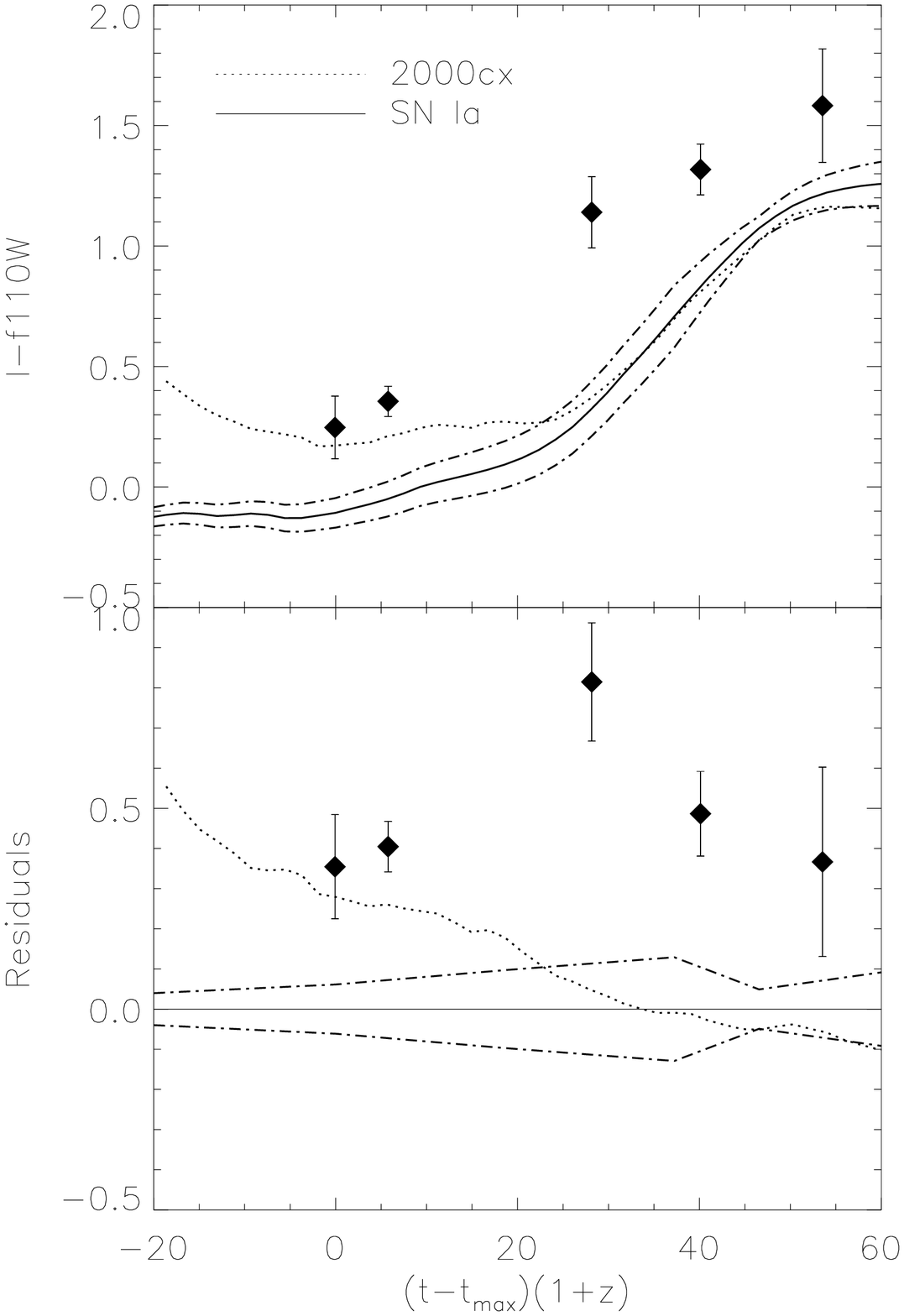}
  \caption{Observer-frame color for SN~1997ek (z=0.863). The color
  curves are obtained by synthetic photometry of the
  spectral templates warped to match the intrinsic colors of SNe~Ia
  (solid line) and  warped to match the colors of the peculiar
  SN~2000cx (dotted line). See text for details. The open symbol
  has not been used for computing the color excess.}
  \label{1997ek}
\end{figure}

Figures~\ref{colors}-~\ref{1997ek} show the 
comparison between the color curves obtained by synthetic photometry
and the observed data. The filled symbols show the observed colors, whereas the
color curves (solid lines) are the expected color evolution. The 
dashed lines represent the intrinsic dispersion for the given color, 
determined using the results in \citet{2008A&A...487...19N}. 
Table~\ref{table:SNe} summarizes the results of this procedure. We note
that the values of the stretch $s$ and color excess $E(B-V)$ are
slightly different from those reported in
\citet{2003ApJ...598..102K}. The discrepancies depend mainly on the
different intrinsic colors assumed, in particular in $U-B$, and
on the spectral template and color-stretch relation used. Since the 
models assumed here are based on a larger sample, small discrepancies
are not surprising (see also the discussion
in \citet{2008A&A...487...19N}). We note that most of the
differences are within one standard deviation from the quoted
uncertainties.

We measured a large color excess, $E(B-V)=0.19 \pm 0.04$, for the lowest
redshift SN in our sample (SN~1998as at $z=0.356$). As shown in
Figure \ref{filters}, the F110W
filter at this redshift covers rest-frame $R$, $I$, and extends
to about 11000~\AA\, effectively covering rest-frame $z$-band.
Thus, the infrared data explore the red region of this SN
spectrum which is less affected by reddening.
These data offer the possibility to explore intrinsic color properties
of SNe~Ia despite the large color excess in the blue bands.  Since
the analysis presented by \citep{2008A&A...487...19N} extends only to
the $I$-band, the colors of the spectral template beyond this band
are obtained by extrapolation to longer wavelengths.
However, the spectral templates are quite 
featureless at these wavelengths and we expect very smooth behavior from 
the $I$-band to the $z$-band.  This gives us confidence about the
average color evolution within the assumed intrinsic dispersion. We
note that the data agree with the expected  color curves.

SN~1997ek shows an interesting color behavior. Its optical color is
slightly bluer than expected and consistent with no extinction, but
its infrared colors are substantially redder than average, as shown in
Figure \ref{1997ek}.  We note that mis-measurement of the SN flux due
to host galaxy contamination can be excluded due to the smooth profile
and low flux level seen in the reference image of the galaxy.
Moreover, the time evolution does not follow the behavior expected
around 20 days after maximum brightness, as shown in the left panel of
Figure \ref{1997ek}. For this reason the color excess has been
estimated using only the filled symbols in the figure.  Given the
different behaviors shown by this SN, we have compared its colors with
those of the very peculiar nearby SN~2000cx
\citep{2001PASP..113.1178L,2003PASP..115..277C,2004ApJ...606..413B}.
Pre-maximum spectra of SN~2000cx resemble that of the over-luminous
SN~1991T and its color evolution is quite different from that of a
normal Type~Ia. The fact that the pre-maximum spectrum of SN~1997ek is
well fitted by spectroscopically SN~1991T-like SN~1999aa, together
with its peculiar colors, has prompted us to attempt a comparison of
these two SNe.  We have warped the spectral template with $UBVRI$
photometry of SN~2000cx before computing synthetic photometry with the
passband filters used for the observations of SN~1997ek. The resulting
color curves are shown as dotted lines in Figure \ref{1997ek}. The
agreement is improved in the infrared around maximum, but not at later
times where SN~2000cx has colors similar to those of a normal SN~Ia.
Moreover, the dotted curve does not fit the observed data in the
optical.  We note that, if only the optical data at maximum were
available, one would conclude that this SN is quite normal.  The
peculiar color behavior of SN~1997ek could be an indication of an
intrinsic peculiarity of this SN. We note that the confidence index
for the spectral typing of SN~1997ek is 4. Although it is not as
secure as candidate that has clear SNe Ia features, such as Si II or S
II, it is unlikely to be a SN of a different type.

The value of $E(B-V)$ determined from the optical data gives a good
agreement between the model and observed infrared color, giving a
value of the $\chisq \sim dof$ for all cases except SN~1997ek.
We take the good agreement found in the case of
SN~1998ay as a confirmation that this is indeed a Type~Ia SN.  

\begin{figure}
  \includegraphics[width=8cm]{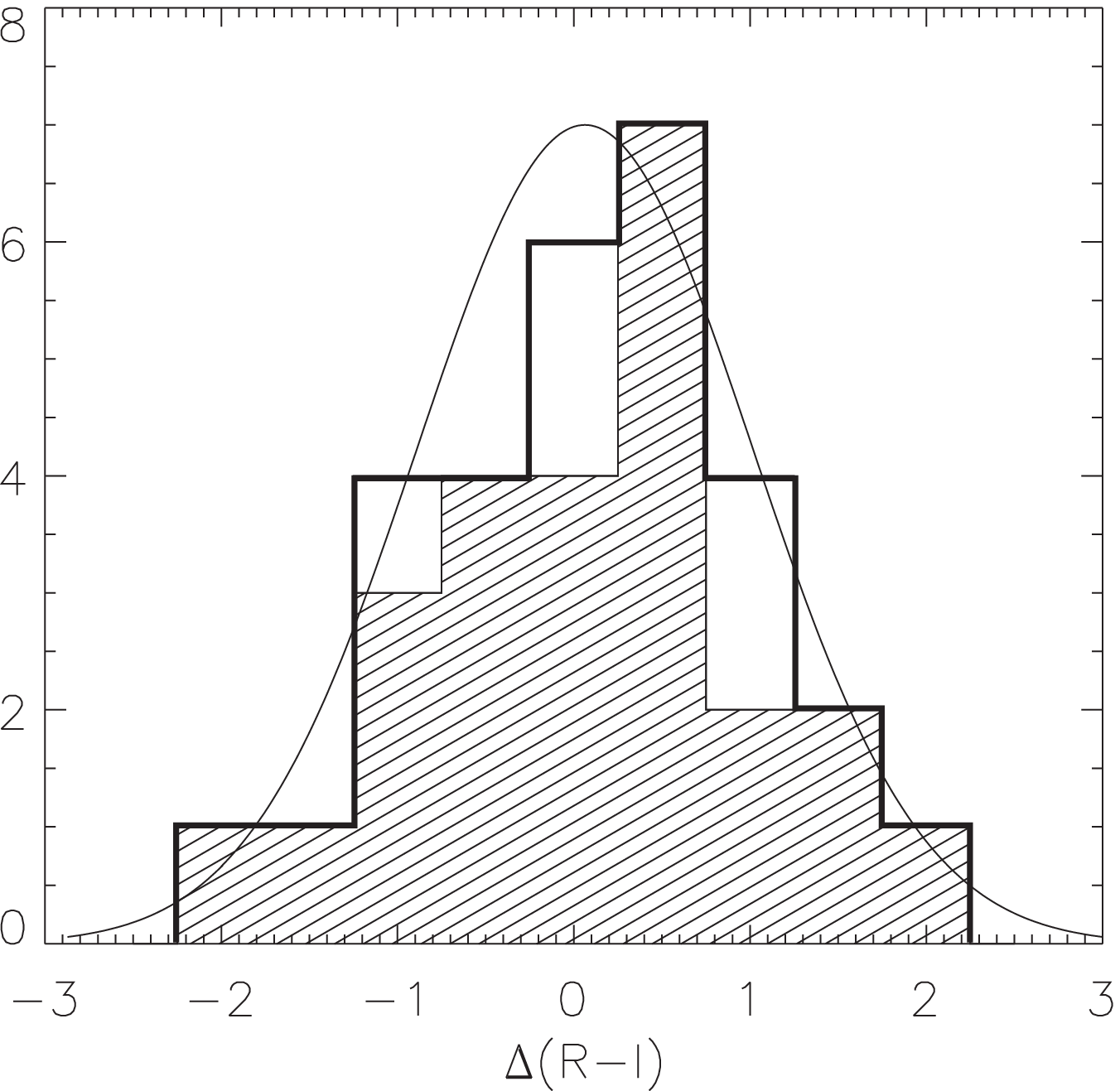}
  \includegraphics[width=8.35cm]{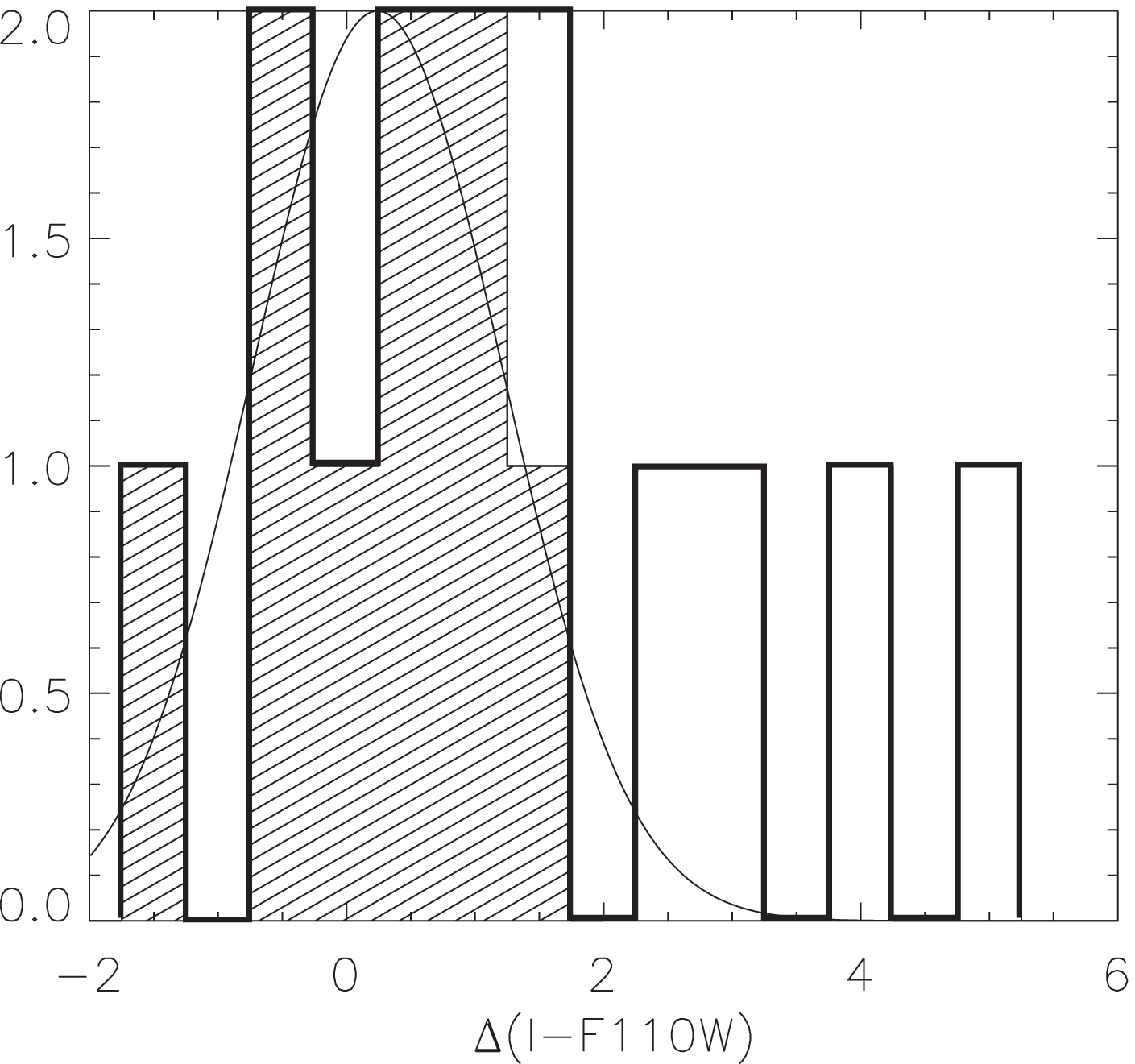}
  \caption{Histograms of the residuals of all data from the expected
     color curves (solid line) and for all SNe but
     SN~1997ek (shaded area). For the latter we measure $\sigma=0.95$,
     for $N=25$ (left panel) in the optical; and $\sigma=0.97$, for
     $N=9$ (right panel) in the infrared. See text for details.}
  \label{histo}
\end{figure}

We compare the total distribution of the residuals of our
sample with the expected distribution of nearby SNe. 
We normalize the residual of each data point from the expected color
by dividing by the total uncertainty.  The total uncertainty
is calculated as a quadratic
sum of the measurement uncertainty and the intrinsic dispersion:
$$\sigma_{tot}=\sqrt{\sigma^2+\sigma_{intr}^2}$$
Figure~\ref{histo} shows the pull distribution for all SNe in each
observed color. The shaded area include 
all SNe except the peculiar SN~1997ek. This is consistent with a 
Gaussian distribution, with zero mean and $\sigma=1$.
Since the intrinsic dispersion used is based on measurements of nearby
SNe, we can conclude that the same average color dispersion is
observed in the distant sample.

\subsection{SN~1999ff and SN~2000fr}

\begin{figure}
  \includegraphics[width=8cm]{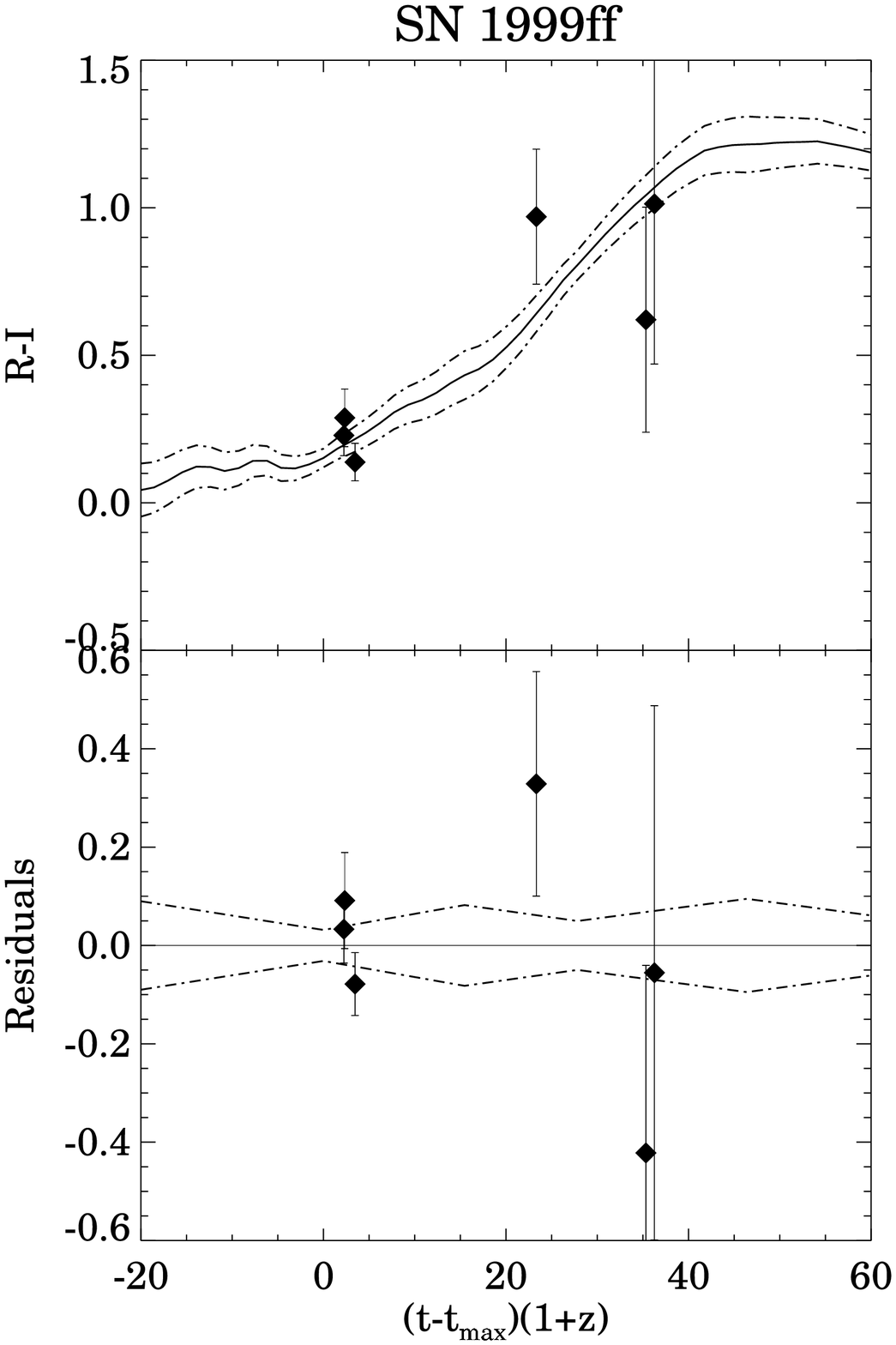}
  \includegraphics[width=8cm]{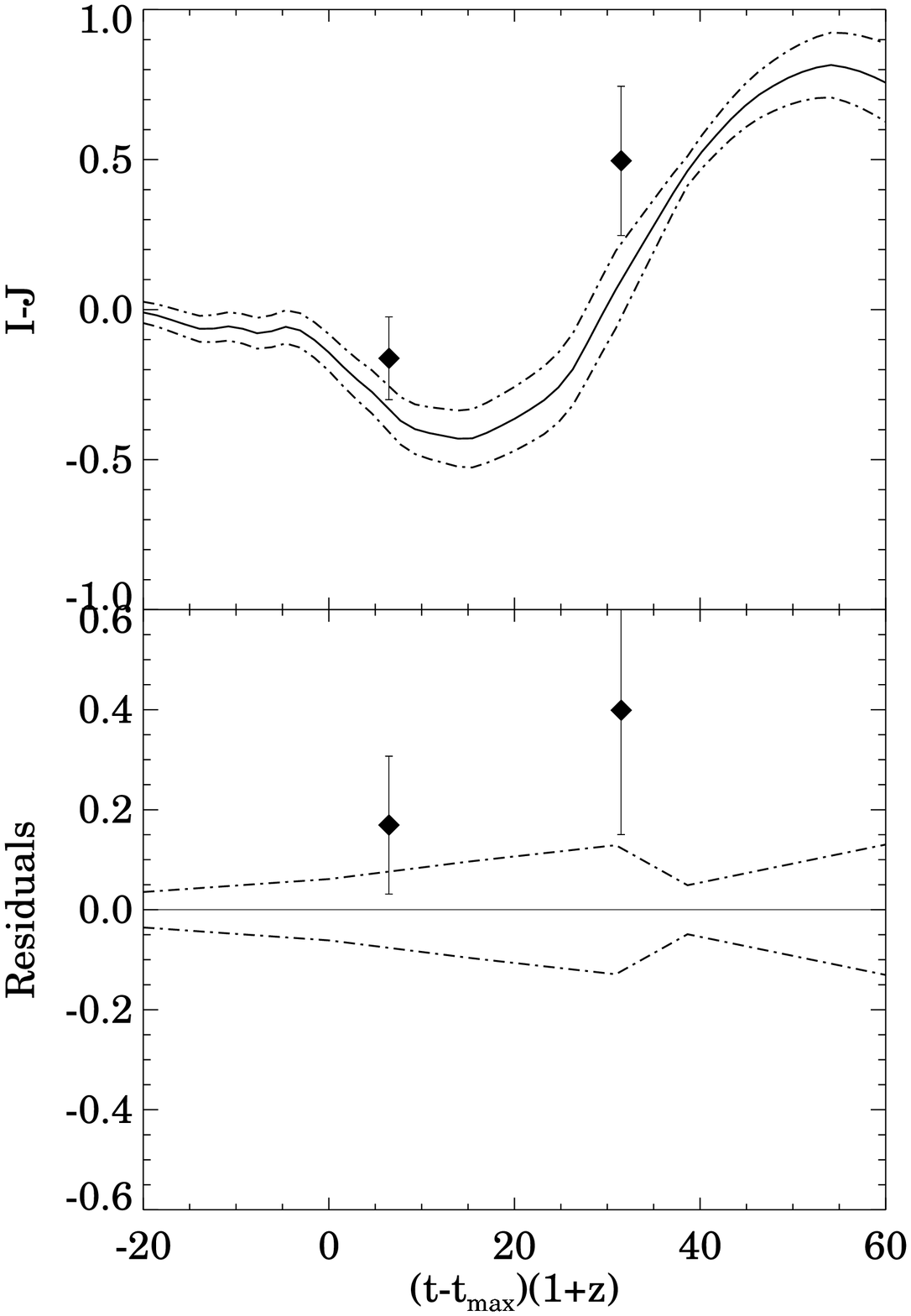}
  \includegraphics[width=8cm]{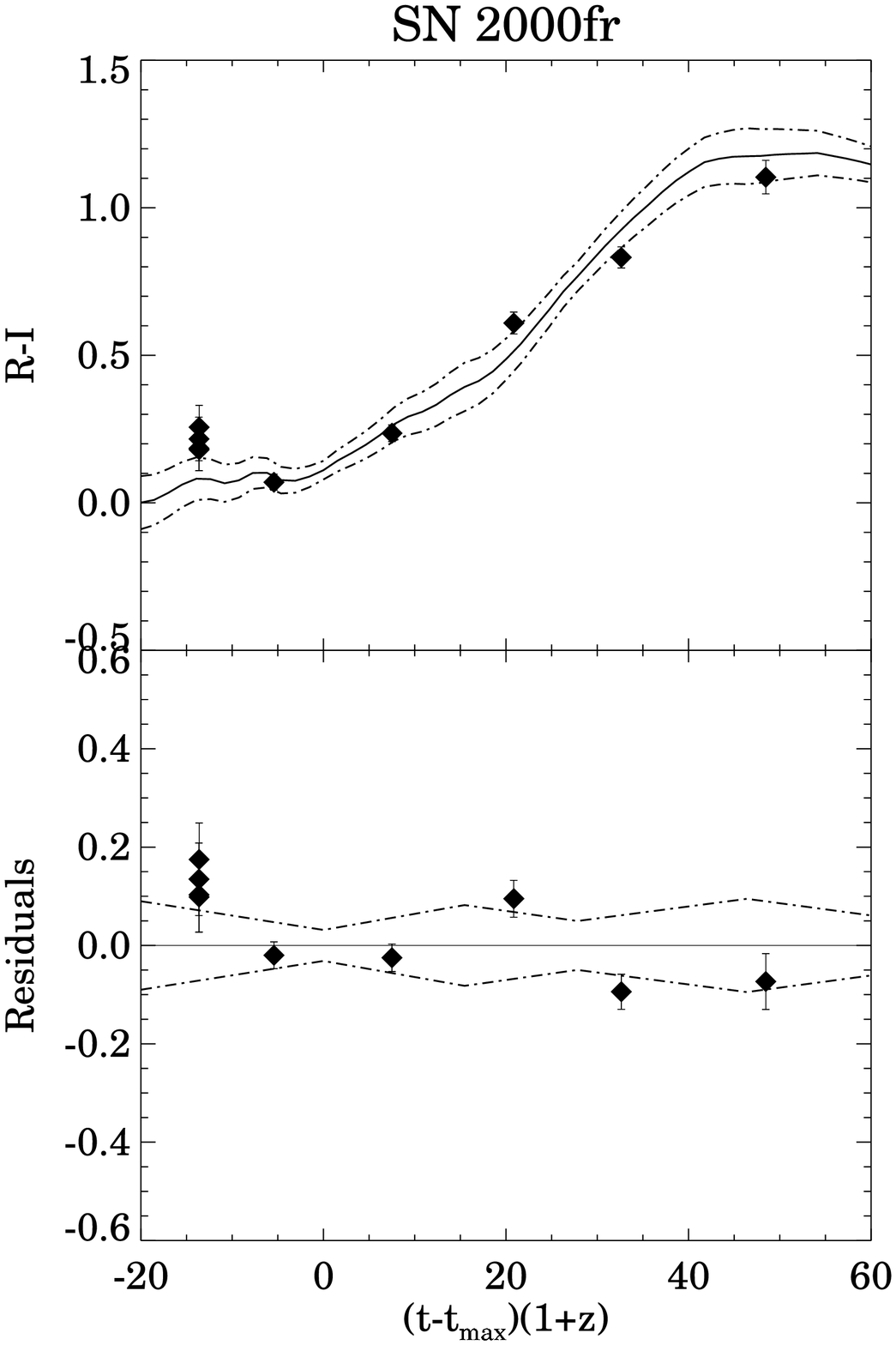} 
  \includegraphics[width=8cm]{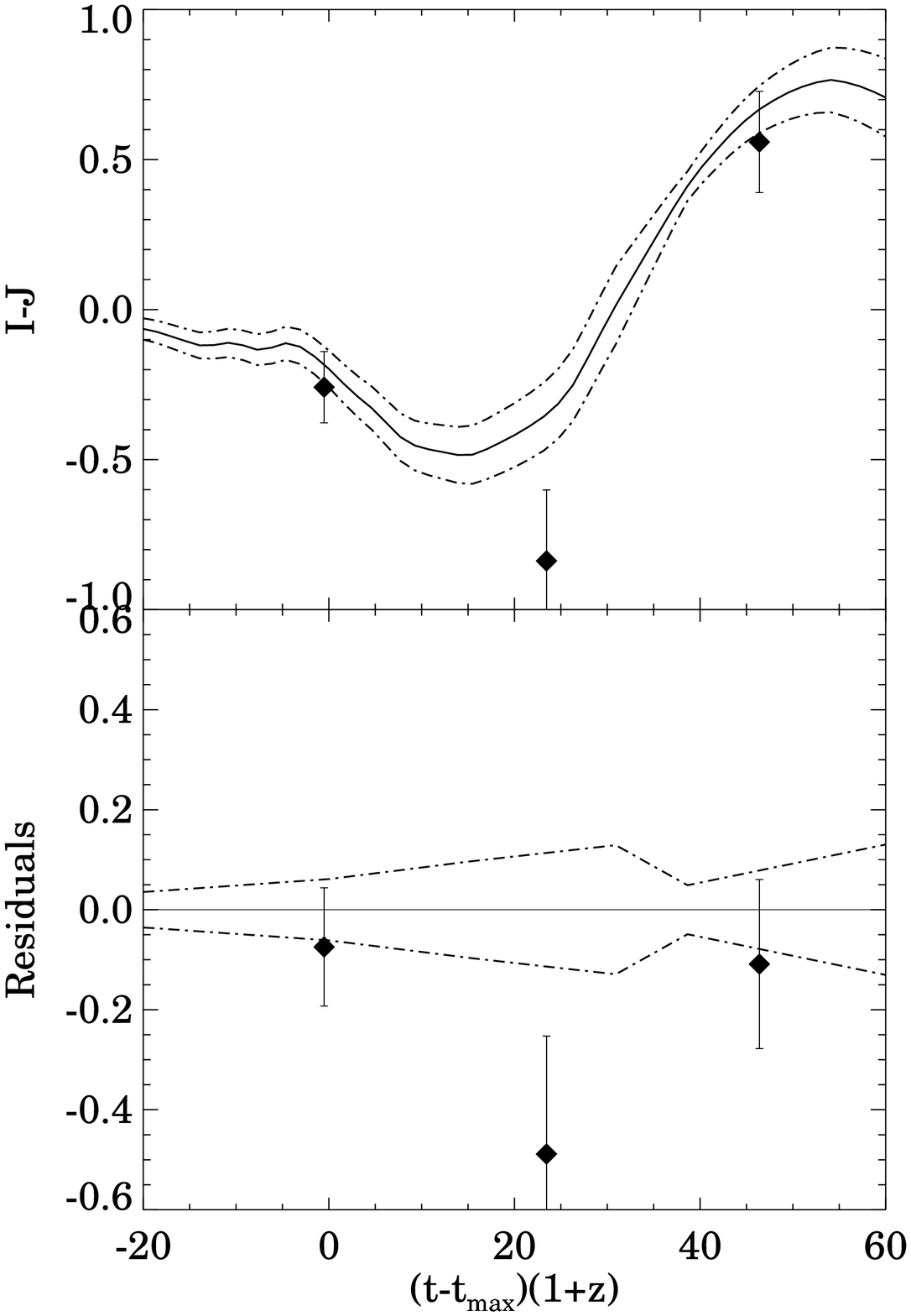}
  \caption{Observed-frame color for SN~1999ff \citep{2003ApJ...594....1T} and SN~2000fr
  \citep{2003ApJ...598..102K,2005A&A...437..789N}. The color curves are 
  obtained by synthetic photometry of the spectral templates (see text
  for details). The dashed lines show the intrinsic dispersion around
  the color curves measured on nearby SNe~Ia
  \citep{2008A&A...487...19N}.}
  \label{colors2}
\end{figure}

\citet{2005A&A...437..789N} presented ground-based infrared
observations for one SN~Ia, SN~2000fr at redshift $z=0.543$. This SN
together with SN~1999ff at $z=0.455$ \citep{2003ApJ...594....1T}
was compared to nearby SNe~Ia and used
to build a rest-frame $I$-band Hubble diagram.
We present the result of a re-analysis of these data, using the same
technique applied to our 5 SNe in the previous
section. Figure~\ref{colors2} shows optical and infrared color
evolutions for both SNe. Table~\ref{table:hizSNe} summarizes the
results.

\begin{deluxetable}{ccccr}
\tablecolumns{5} 
\tablewidth{0pc}
\tablecaption{Parameters of the Additional SNe, Redshift, $B$-band Stretch Factor,  Milky Way Reddening, and Color Excess.} 
\tablehead{SN\tablenotemark{a} & $z$ & $s_B$ & $E(B-V)_{MW}\tablenotemark{b}$ & $E(B-V)$}
\startdata
SN~1999ff  & 0.455  & 0.82 $\pm$ 0.05 & 0.025 & $0.06 \pm 0.11$\\
SN~2000fr  & 0.543  & 1.054 $\pm$ 0.012 & 0.030 & $0.01 \pm 0.04$\\
\enddata 
\tablenotetext{a}{From \citet{2005A&A...437..789N}}
\tablenotetext{b}{Given by \cite{1998ApJ...500..525S}}
\label{table:hizSNe}
\end{deluxetable}

We note that the color excess giving the best fit to the optical
data is consistent with no reddening for both SNe. The infrared data also agree
with the expected behavior within the uncertainties.  

\section{$I$-band Hubble diagram}
\label{sec:HD}

Two SNe in our sample are at a redshift where the infrared
observations correspond to the rest-frame $I$-band (see 
Figure~\ref{filters}). These are SN~1998as ($z=0.356$) and SN~1998ay 
($z=0.638$). We note that SN~1998as has the largest extinction in our sample and SN~1998ay 
does not have a confirming spectrum, but is fully
consistent with a SN~Ia due to its colors and light curve shapes.
We used these SNe to extend the $I$-band Hubble diagram
built by \citep{2005A&A...437..789N} which originally included 2 $z>0.1$ SNe:
SN~2000fr at $z=0.543$ and SN~1999ff at $z=0.455$.

\begin{figure}
  \includegraphics[width=8cm]{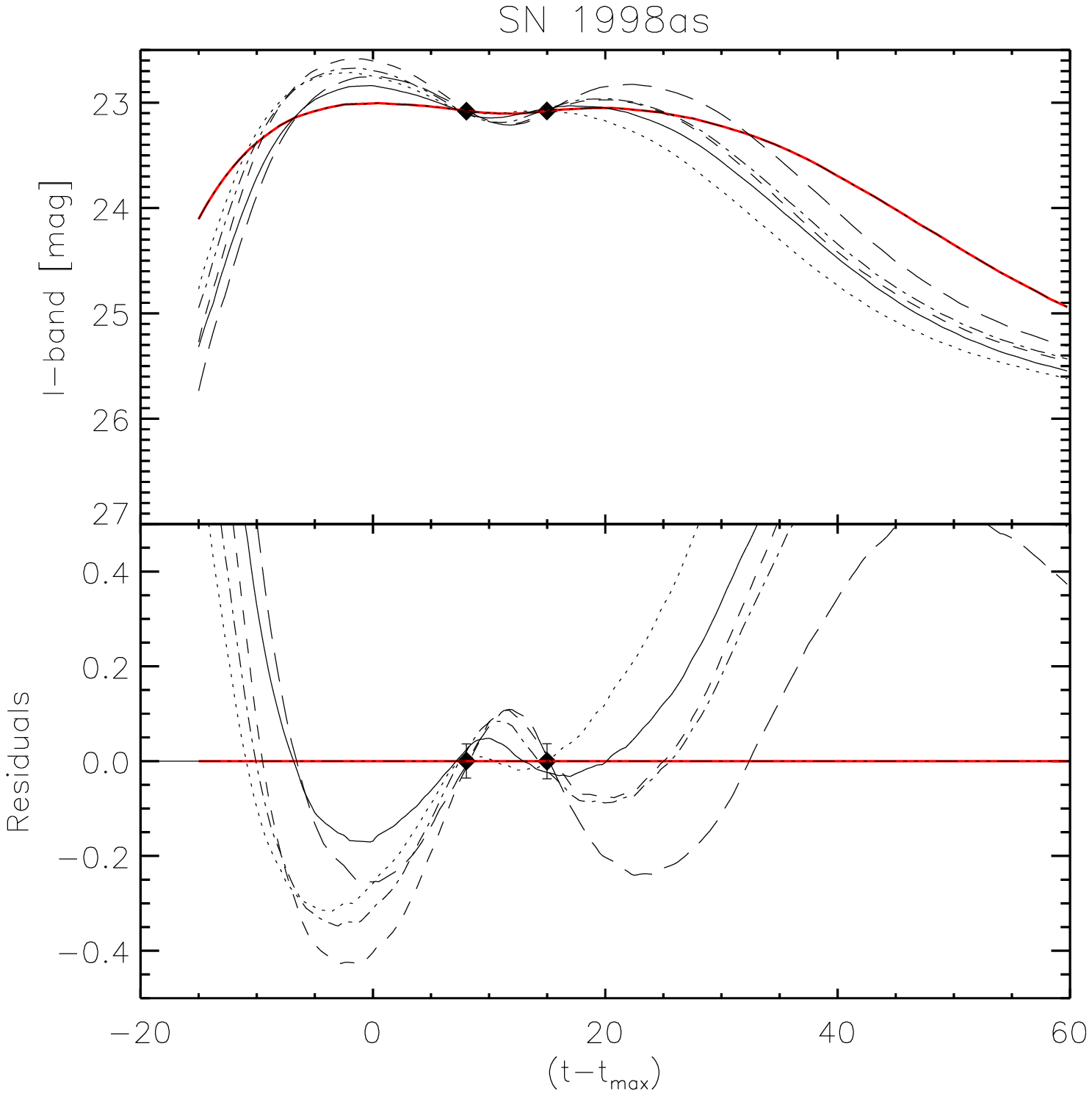}
  \includegraphics[width=8cm]{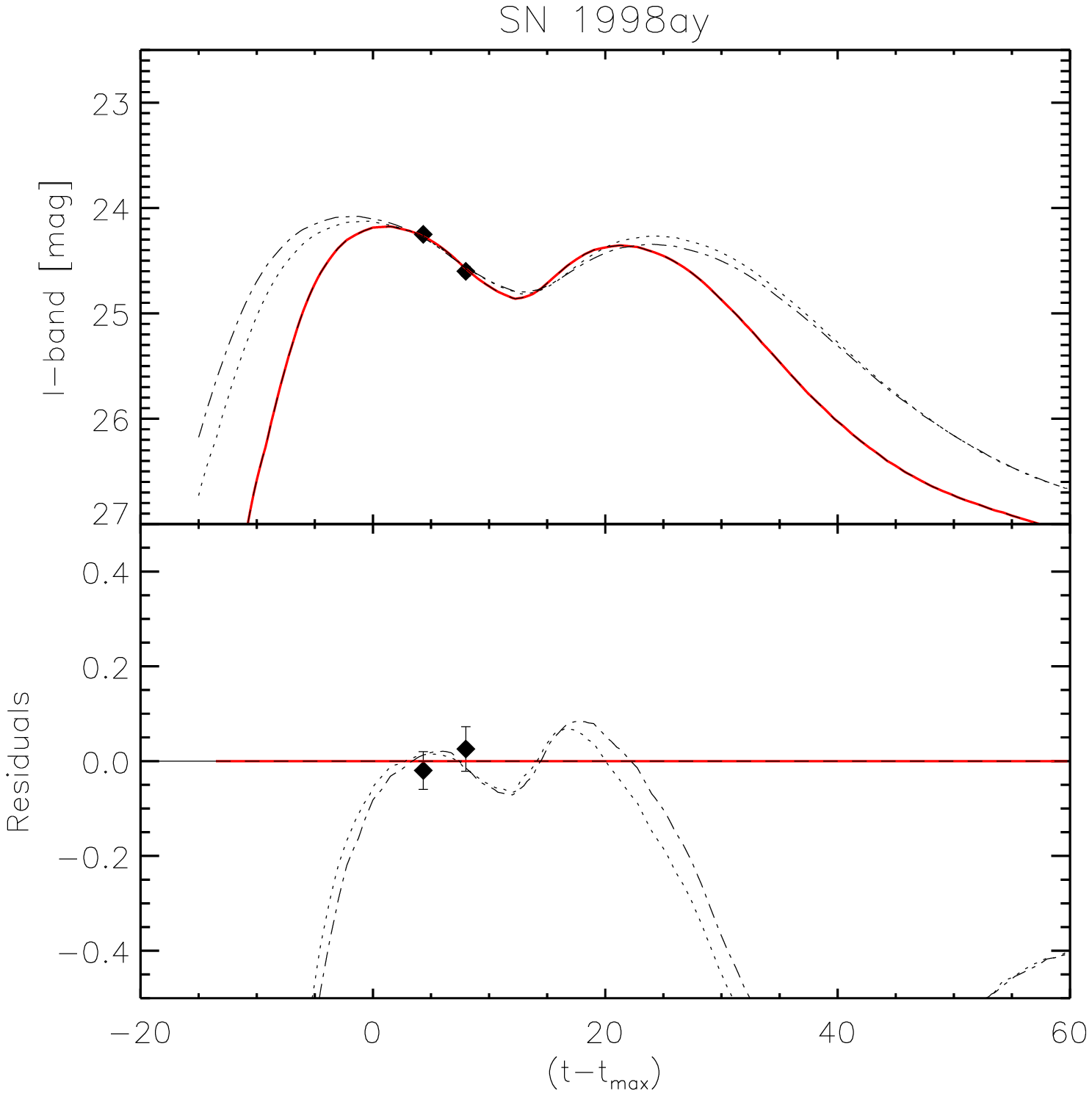}
  \includegraphics[width=8cm]{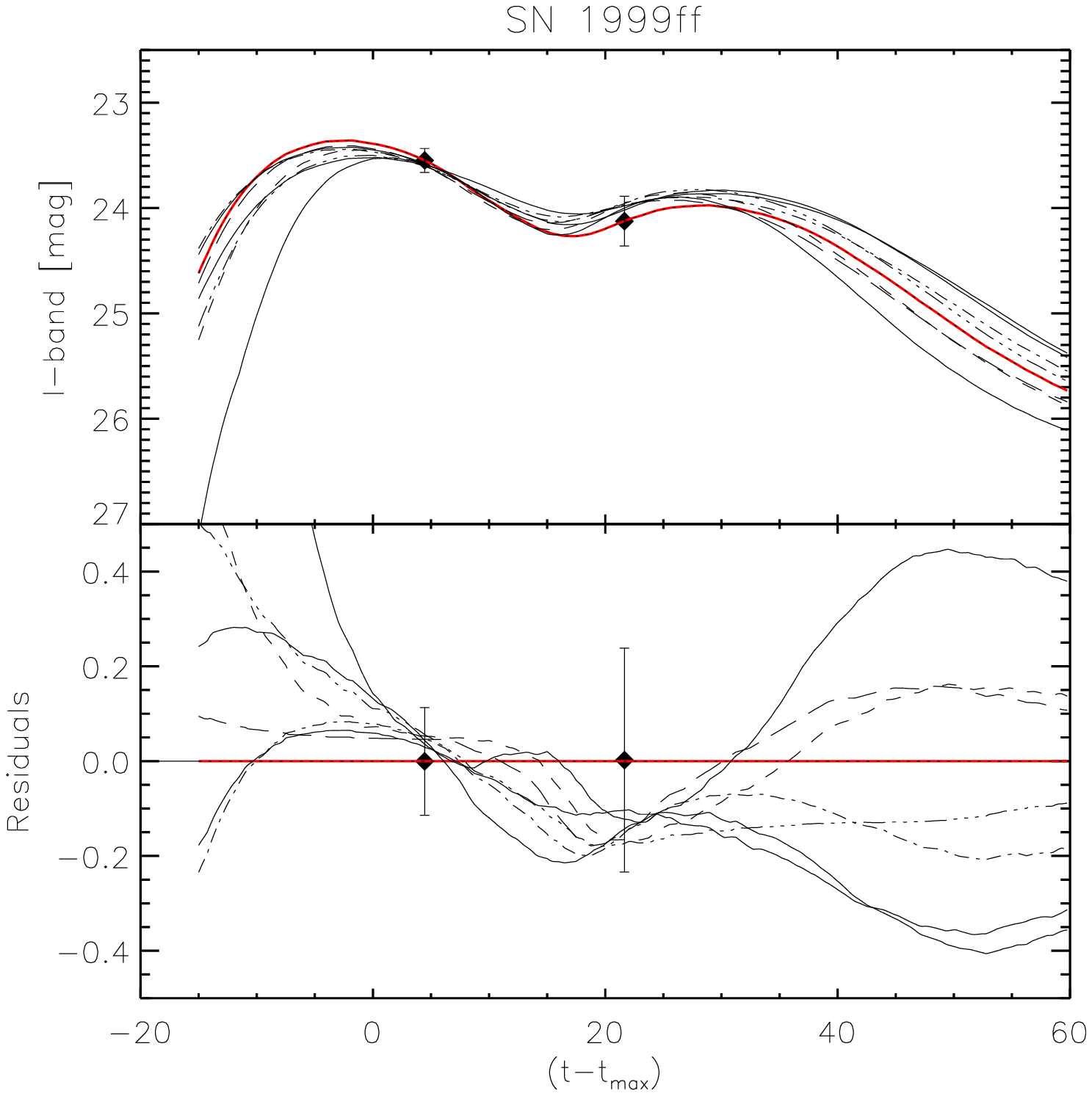}
  \includegraphics[width=8cm]{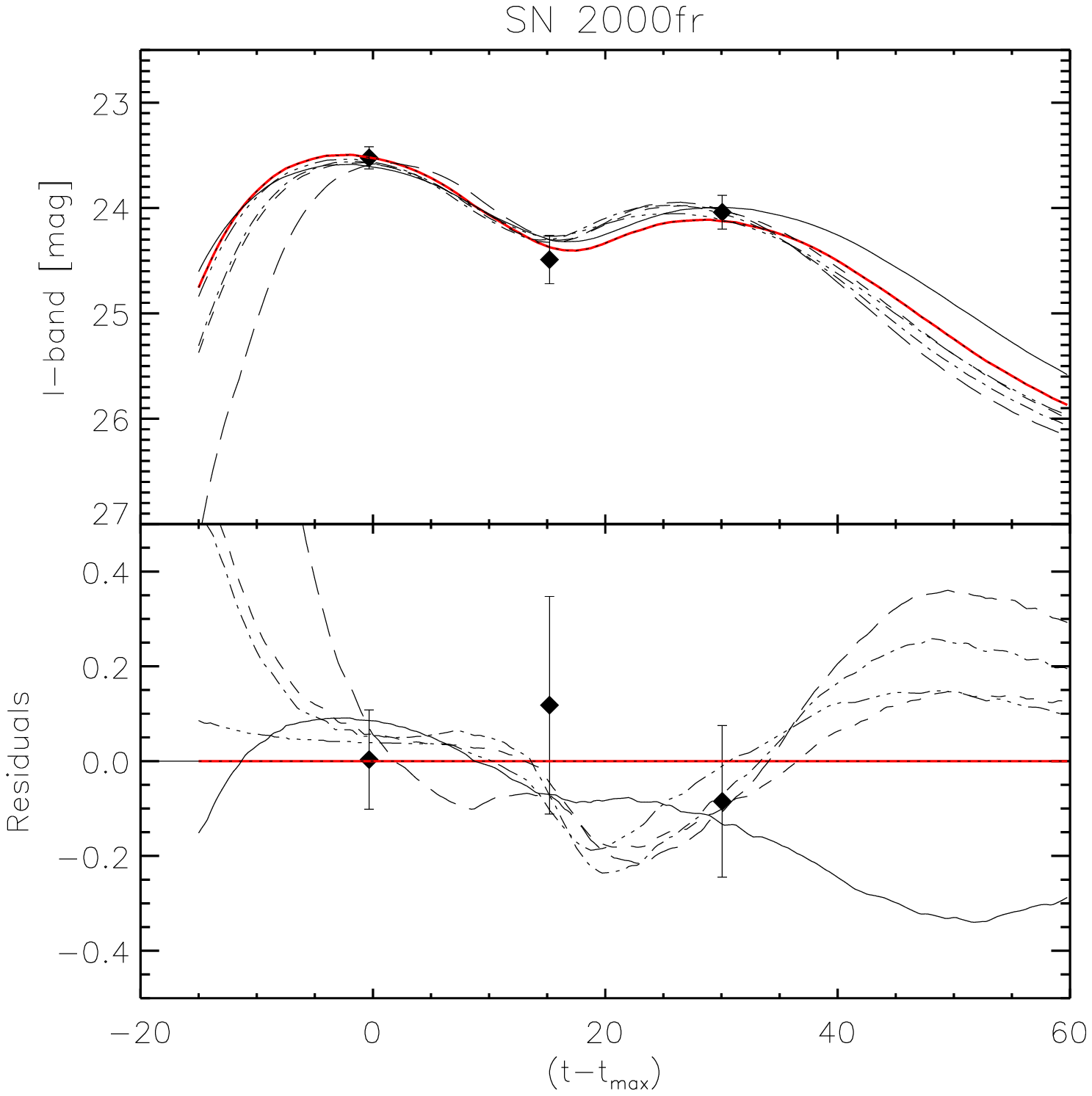}
  \caption{Best fits to rest-frame $I$-band observations for  
  SN~1998as, SN~1998ay, SN~1999ff and SN~2000fr. All templates giving
  a $\chisq < \chisq_{min}+1$ are shown together with the best fit template
  (thicker red solid line). The best fit templates are listed in Table~\ref{tb:iband}}
  \label{iband}
\end{figure}

We computed $K$-corrections for all four SNe from the observed
infrared filter to rest-frame $I$-band following
\citet{2008A&A...487...19N}. Using the $I$-band templates of 42
nearby SNe~Ia developed in \citet{2005A&A...437..789N} we fit the
peak brightness of the high-redshift SNe. Figure~\ref{iband} shows 
the template fits that satisfied $\chisq \le \chisq_{min}+1$, where
$\chisq_{min}$ is the $\chisq$ of the best fitting template. The
fitted $I_{max}$ giving the minimum $\chisq$ is reported in
Table~\ref{tb:iband}. However, since several templates correspond to a
similar value of the $\chisq$, we estimated the uncertainty, $\Delta$, 
as the rms of the peak magnitudes that satisfied $\chisq \le
\chisq_{min}+1$. 

Note that, given the different way of computing $K$-corrections, 
the results obtained for SN~2000fr and SN~1999ff are slightly
different than what was found in \citet{2005A&A...437..789N}.

\begin{deluxetable}{ccccccc}
\tablecolumns{4} 
\tablewidth{0pc}
\tablecaption{Results of the Rest-Frame $I$-band Light Curve Fits for Two
of the SNe in our Sample, Together with SN~2000fr and SN~1999ff.}
\tablehead{SN & $z$ & $I_{max}$\tablenotemark{a}  &
  $\Delta$\tablenotemark{b} & Best Template & Templates\tablenotemark{c}}
\startdata 
SN~1998as &  0.356 &  23.004    &   0.146 &   1998ab &  1993H,1994D,1997E,1998V,1999cl\\
SN~1998ay &  0.638 &  24.175    &   0.048 &    1994T &  1992bp,1997dg\\
SN~2000fr &  0.543 &  23.495    &   0.033 &   1992bc &  1992bh,1994ae,1996bl,1999gp\\
SN~1999ff &  0.455 &  23.359    &   0.058 &   1992bc & 1992bh,1999gp,1996bl,1999aa,\\
          &        &            &         &          &  1996C,1995D,1992bh\\
\enddata
\tablenotetext{a}{Peak magnitude obtained by the best fit
  template.}
\tablenotetext{b}{The rms magnitude of the best-fit templates that satisfied $\chisq < \chisq_{min}+1$. See text for
  details.} 
\tablenotetext{c}{Best fit templates giving a $\chisq <
  \chisq_{min}+1$. See text for details.}
\label{tb:iband}
\end{deluxetable}

The peak magnitudes have been corrected for the width-luminosity
relationship in the $I$-band and the Milky Way absorption,
$A^{MW}$. The host galaxy extinction correction, $A_I^{host}$, was
estimated from the measured color excess and assuming the CCM
extinction law with $R_V=1.75$:

\begin{equation}
m_I^{\rm eff}=m_I+\alpha_I(s_B-1)-A_I^{host}-A^{MW}
\label{Imax}
\end{equation}

\begin{figure}
  \includegraphics[width=15cm]{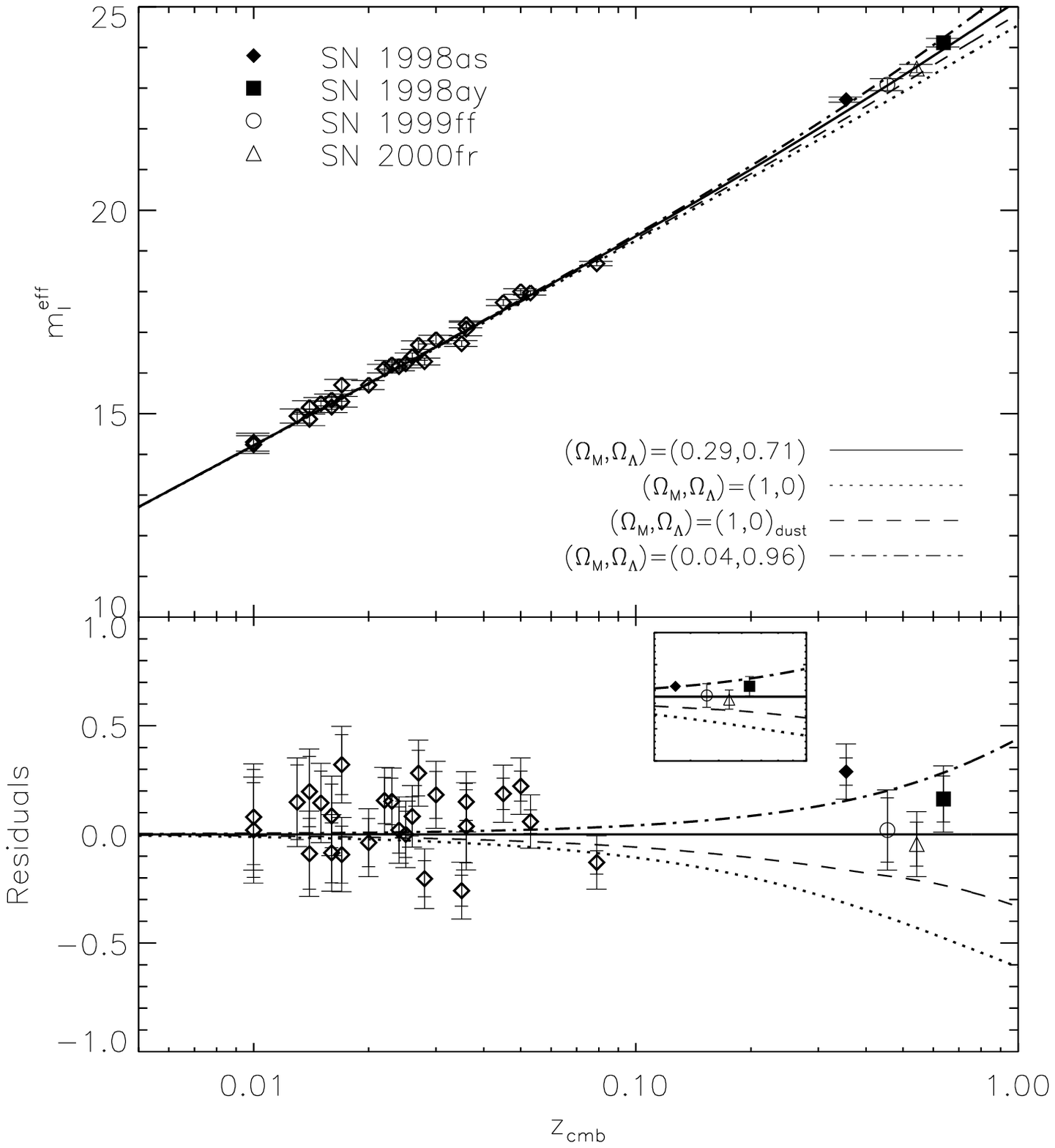}
\caption{Rest-frame $I$-band Hubble diagram. The open symbols are data
  from \citet{2005A&A...437..789N}. The solid symbols are 2 high-z SNe
  from this work. The solid line is the best fit to the nearby data
  for the concordance model with fixed $\Omega_M=0.29$ and
  $\Omega_\Lambda=0.71$. The dotted line is a flat universe model with
  no cosmological constant, while the dashed line is the same model in
  presence of large grain dust in the intergalactic medium. The
  dashed-dotted line is a flat universe with no dark matter,
  $\Omega_M=0.04$, and $\Omega_\Lambda=0.96$. The inner error bars
  represent the uncertainties reported in Table~\ref{tb:iband}, while
  the outer error bars are obtained adding an intrinsic dispersion
  equal to 0.11 mag.  The inset shows the effect of using $R_V = 3.1$
  for host galaxy dust extinction.}

\label{HD}
\end{figure}

\noindent The effective magnitudes of the four high redshift SNe
together with the effective magnitudes of nearby SNe
\citep{2005A&A...437..789N} have been used to build the rest-frame
$I$-band Hubble diagram shown in Figure~\ref{HD}. A value of
$\alpha_I=1.18 \pm 0.19$ was fitted for the low-redshift sample and
used for all SNe.  The uncertainties reported in Table~\ref{tb:iband}
are represented by the inner error bars, and the larger error bars are
obtained by adding an intrinsic dispersion equal to 0.11 mag as
measured on the nearby SNe.  The rms dispersion of the
high-redshift SNe is 0.15 mag.

The solid line represents the best fit to the rest-frame $I$-band
nearby data for the concordance model with fixed $\Omega_M=0.29$ and
$\Omega_\Lambda=0.71$, as found by \citet{2008ApJ...686..749K} in the
$B$-band. The single fitted parameter, $\mathcal M_I$, is defined as
in \citet{1997ApJ...483..565P}:

\begin{equation}
{\mathcal M}_I\equiv M_I-5\log H_0+25
\end{equation}

\noindent where $M_I$ is the $I$-band absolute magnitude for a
$B$-band stretch $s_B=1$ supernova. The value fitted is $\mathcal M_I
=-3.19 \pm 0.02$. 
The dotted line in Figure~\ref{HD} represents a cosmological model without 
dark energy ($\Omega_M=1$ and $\Omega_\Lambda=0$) while the dashed-dotted line
represents a cosmological model without dark matter ($\Omega_b=0.04$ and
$\Omega_\Lambda=0.96$), where the matter density is equal to the baryon
matter density only. Also plotted is the model of a universe without dark  
energy in the presence of a homogeneous distribution of large grain
dust in the intergalactic medium (dashed line). 
\citet{2002A&A...384....1G} hypothesized an ad-hoc
distribution of dust that is able to explain the dimming observed in
the SNe~Ia $B$-band peak magnitudes. According to this model, the comoving 
density of the dust increases with cosmic time until z=0.5.
This ``replenishing dust'' is created at the same rate as universal expansion.
By construction it is very difficult to constrain this kind of dust
model using the $B$-band Hubble diagram alone \citep[see
  e.g.,][]{2007ApJ...659...98R}. We are able to exclude this model
with high confidence with the rest-frame $I$-band Hubble diagram, as
shown in Table~\ref{testcosmo}. All values of $\chisq$ computed with the
four high-z SNe for each model tested are also reported in the table. 

To see the effect of different values of $R_V$ on the Hubble diagram,
we repeated the analysis in the case of a standard Milky Way type dust
also in the host galaxy, with $R_V=3.1$. The results are summarized  in 
Table~\ref{testcosmo}. It is worth noticing that the constraint put on the 
replenishing dust model are not equally stringent in this case. The residuals 
of the high redshift SNe are shown in the inset in Figure~\ref{HD}. We note that
the dispersion of the data set is smaller in this particular case for $R_V=3.1$ 
than for $R_V=1.75$.

We repeat the cautionary note in \citet{2005A&A...437..789N}.
The different method applied to determine the peak magnitudes for the
nearby and distant SNe~Ia might introduce systematic
uncertainties which are difficult to quantify. Moreover, many (but not
all) of the templates used for fitting the high-redshift SN are built
on the nearby supernovae in the Hubble diagram leading to a possible
source of correlation. 

\begin{deluxetable}{lrc|rc}
\tablecolumns{3} 
\tablewidth{0pc}
\tablecaption{$\chisq$ with 4 d.o.f. for the Cosmological Models in Figure \ref{HD}} 
\tablehead{ $(\Omega_M,\Omega_{\Lambda})$ & $\chisq$\tablenotemark{a} & $P(\chisq < \chisq_0)$\tablenotemark{a} & $\chisq$\tablenotemark{b} & $P(\chisq < \chisq_0)$\tablenotemark{b}}
\startdata 
(0.29,0.71)     & 7.04  & 0.87  & 0.78  & 0.06\\
(0.04,0.96)     & 6,04  & 0.80  & 6.21  & 0.82\\
(1,0)           & 54.06 & 1.00  & 25.71  & 1.00\\
$(1,0)_{dust}$  & 23.64 & 1.00  & 7.75  & 0.90 \\
\enddata
\tablenotetext{a}{Results obtained assuming $R_V=1.75$}
\tablenotetext{b}{Results obtained assuming $R_V=3.1$}
\label{testcosmo}
\end{deluxetable}

\section{Summary and discussion}
\label{sec:discussion}

This data set is unique in terms of wavelength coverage and
redshift. In fact, only a handful of SNe~Ia at these redshifts have
multi-epoch observations in the near-IR.
\citet{2007ApJ...659...98R} presented NICMOS observations
of SNe~Ia at $z > 1$. However, the majority of the data
correspond to rest-frame $B$ and $V$ and are
mostly at later epochs between 5 and 20 days after maximum brightness.

In Section \ref{sec:spectra}, we presented the spectra that were used
for SN typing and for measuring the redshift. All SNe were securely
identified as Type Ia except SN~1998ay which has a spectrum suffering from
severe host galaxy contamination. SN~1998ay shows all other
characteristics of a Type~Ia SN. The observed colors are in 
perfect agreement with the models in the optical and in
the infrared, providing strong evidence that this is a Type~Ia SN. 

The high quality of the infrared data allowed a detailed comparison of 
the observed optical-IR colors of these high-redshift SNe~Ia
to the colors of nearby SNe~Ia. This was presented in 
Section \ref{sec:colour}. Optical colors were used to determine
the color excess in rest-frame $E(B-V)$. The color excess correction
to be applied to the infrared was determined assuming
the CCM extinction law. 
We note that the small amount of absorption in the infrared compared to the
optical bands and the relatively large uncertainties in the measurements
undermine the
possibility of determining $R_V$. This is even more severe for those
SNe with small color excesses such as
SN~1998ay. For this reason we have assumed a nominal standard value of
$R_V=1.75$ as in \citet{2008A&A...487...19N}. Assuming different
values of $R_V$, including the commonly assumed value of 3.1,
lead to estimates with similar color-excess. We note a 
generally good agreement with the expected color evolution in both 
$R-I$ and in $I-F110W$.  

SN~1997ek is the most distant SN in our sample and the only SN with
unusual colors. It is found to have a normal optical color around
maximum light corresponding to rest-frame $U-B$.  However, it is
significantly redder than average in the observer-frame $I-F110W$,
corresponding to rest-frame $B-VReI$, where $VReI$ is a combination of
$V$, $R$, and a small fraction of $I$. This behavior is not compatible
with a reddening process which by definition would affect shorter
wavelengths more than longer ones. Thus, we take it as an indication
of a peculiarity in the intrinsic color of this SN. We found no
counterpart of this object in the nearby sample analyzed by
\citet{2008A&A...487...19N}.

The spectrum of SN~1997ek at 3 days pre-maximum is
similar to that of the peculiar SN~1999aa, which is spectroscopically
similar to the over-luminous SN~1991T \citep{2004AJ....128..387G}. 
Including all 1991T-like, 1991bg-like and 1986G-like SNe, about 36\% of all
SNe~Ia in the nearby sample are peculiar SNe~Ia
\citep{2001ApJ...546..734L}.  However, only a small fraction of peculiar 
SNe~Ia are discovered at high redshift
\citep[see
  e.g.,][]{2005AJ....129.2352M,2006AJ....131.1648B, 2007A&A...470..411G,2008A&A...477..717B}.
On the other hand wholly unknown SN subtypes are discovered at high-z,
e.g., SNLS-03D3bb \citep{2006Natur.443..308H} and SCP 06F6
\citep{2009ApJ...690.1358B}. 
The different rate of peculiar objects discovered could be taken as an
indication of evolution of SN properties with redshift. However, other
explanations are also possible.
For instance, magnitude-limited surveys could fail
to detect sub-luminous SNe or over-luminous SNe suffering high dust
extinction. The lower signal-to-noise ratio in the observed
spectra of high-z SNe~Ia also limits the possibility of discovering
spectroscopic peculiarities \citep[see discussion
  in][]{2001ApJ...546..734L}. Multi band photometry offers a very 
useful tool for discriminating between SN subtypes by allowing a comparison
of colors which can be measured with higher signal-to-noise
ratios. Identifying properties that distant SNe have in common
with nearby SNe, even if peculiar, supports the validity
of SNe~Ia as distance indicators to probe the
composition of the universe. Future large SNe~Ia datasets with
multi-band light curves (such as the SNLS) will help to determine
whether other SNe like SN~1997ek exist at high-z.

Another intriguing possibility is that this peculiarity is an effect
of varying metallicity. This would have a small effect on the
optical peak brightness of the supernova but would significantly impact the UV
spectral energy distribution
\citep{1998ApJ...495..617H,2000ApJ...530..966L}.
\citet{2008ApJ...674...51E} measured the UV flux in
36 high redshift SNe in the SNLS sample and did not find significant evolution. 
More recently, \citet{2008MNRAS.391.1605S} showed that enhanced Fe and Ti/Cr
abundances result in an increase of flux in the UV. In particular,
an increase of iron abundance affects the region bluer than 3000 \AA\
and leads to a variation in the $U$-band magnitude of up to
0.5-0.7 mag. We note that the $F336W$ pass-band filter
analyzed in Figures 4 and 5 of the \citet{2008MNRAS.391.1605S} analysis
match our observed $R$-band for SN~1997ek. We
tested the hypothesis of increased flux in the $U$-band due to
metallicity for SN~1997ek with Monte Carlo 
simulations.  The spectral template was first modified by an
increment in flux corresponding to a decrement in the blue
end of the spectrum, $\Delta U$.
Color excess corrections to the spectrum were then applied 
for a range of $E(B-V)$ values
with $R_V=1.75$ before computing the synthetic photometry. 
Both optical and infrared data were used to find the best
match to the synthetic color curves: $\Delta
U=0.7^{+0.3}_{-0.2}$ mag and $E(B-V)=0.5^{+0.1}_{-0.2}$ mag with
$\chisq=6.26$ for 8 $dof$.  The model clearly provides
a very good fit to this particular supernova.
Thus, we have utilized one of the models developed by
\citet{2008MNRAS.391.1605S}, for SN~2001ep, giving a similar $\Delta
U$, to modify the colors of SN~1997ek. The model is obtained by 
increasing the iron abundance by +2 dex. 
We obtained the best fit to our data, for $E(B-V)=0.40 \pm {0.05}$
mag, for $\chisq=13.8$ for 9 $dof$.  We note that such value for the color
excess, would imply a large correction to the $B$-band peak magnitude,
that would in turn move this SN about 1 mag away from the best cosmology fit
in the Hubble diagram.

Finally, we used the infrared observations of SN~1998ay and SN~1998as to build
a rest-frame $I$-band Hubble diagram. We have added these two SNe to
the sample from \citet{2005A&A...437..789N} to extend the Hubble
diagram up to $z=0.638$, doubling the high redshift sample. All four
high redshift SNe were treated consistently to fit the $I$-band peak
magnitudes. We found the data to be consistent with a $\Lambda$
dominated flat universe, with ($\om,\;\ola$) = (0.29, 0.71), for both 
values of $R_V$ assumed. 
The smaller dispersion in the $I$-band and the
smaller extinction correction uncertainties allow us to exclude the
presence of a homogeneous distribution of large grain dust in the
intergalactic medium at the five standard deviation level with only 4
SNe~Ia for the case of $R_V=1.75$. This conclusion is severely weakened 
for $R_V=3.1$.
We note, however, that the data set used in this paper is very
inhomogeneous. Moreover, we applied a different method for fitting
the light curve peak magnitude for the nearby and the distant SNe which 
could be a source of systematic uncertainties. 

Multiband observations such as those provided by the
proposed SNAP satellite will be a powerful method for sub-typing of 
SNe~Ia and for determining precise distances.
Observations of high redshift SNe~Ia in the near infrared
will extend the rest-frame $I$-band Hubble diagram
\citep[e.g.,][]{2006AAS...209.9004B}.
Already proven to be an excellent test for systematic 
uncertainties, this approach will continue to compliment the
more extensively used rest-frame SN $B$-band Hubble diagram.

\acknowledgements

We would like to thank Daniel Sauer for sharing his models
with us.
Financial support for this work was provided by NASA through program GO-07850 from
the Space Telescope Science Institute, which is operated by AURA, Inc., under NASA
contract NAS 5-26555.
This work was also supported in part by the Director, Office of Science, Office of
High Energy and Nuclear Physics, of the U.S. Department of Energy under Contract
No. AC02-05CH11231.  
Some of the data presented herein were obtained at the W.M. Keck Observatory, which
is operated as a scientific partnership among the California Institute of
Technology, the University of California, and the National Aeronautics and Space
Administration.
The authors wish to recognize and acknowledge the very significant cultural
role and reverence that the summit of Mauna Kea has always had within the
indigenous Hawaiian community.
We are most fortunate to have the opportunity to conduct observations from
this mountain.
Based in part on observations collected at the ESO La Silla
Observatory (ESO program 60.A-0586).

{\it Facility:} \facility{HST(NICMOS)},\facility{ESO:3.6
  (Optical)},\facility{Keck:II (Optical)} 

\bibliography{../../../bibtex/bib}

\begin{thebibliography}{49}
\expandafter\ifx\csname natexlab\endcsname\relax\def\natexlab#1{#1}\fi

\bibitem[{{Astier} {et~al.}(2006)}]{2006A&A...447...31A} {Astier}, P., et al. 2006, \aap, 447, 31

\bibitem[{{Barbary} {et~al.}(2009)}]{2009ApJ...690.1358B} {Barbary}, K., et al. 2009, \apj, 690, 1358

\bibitem[{{Barker} \& Dahlen(2007)}]{BarkerDahlen}
{Barker}, E., \& Dahlen, T. 2007, NICMOS Instrument Handbook, Version 10.0,
  (Baltimore, MD: STScI)

\bibitem[{{Blondin} {et~al.}(2006){Blondin}}]{2006AJ....131.1648B}
{Blondin}, S., et al. 2006, \aj, 131, 1648

\bibitem[{{Bohlin} {et~al.}(2005){Bohlin}, {Linder}, \& {Riess}}]{2005ISR02}
{Bohlin}, R., {Linder}, D., \& {Riess}, A. 2005, Instrument Science Report
  NICMOS 2005-002, (Baltimore, MD: STScI)

\bibitem[{{Branch} {et~al.}(2004){Branch}}]{2004ApJ...606..413B}
{Branch}, D., et al. 2004, \apj, 606, 413

\bibitem[{{Bronder} {et~al.}(2008){Bronder}, {Hook}, {Astier}, {Balam},
  {Balland}, {Basa}, {Carlberg}, {Conley}, {Fouchez}, {Guy}, {Howell}, {Neill},
  {Pain}, {Perrett}, {Pritchet}, {Regnault}, {Sullivan}, {Baumont}, {Fabbro},
  {Filliol}, {Perlmutter}, \& {Ripoche}}]{2008A&A...477..717B}
{Bronder}, T.~J., et al. 2008, \aap, 477, 717

\bibitem[{{Burns} {et~al.}(2006){Burns}, {Wyatt}, \&
  {Freedman}}]{2006AAS...209.9004B}
{Burns}, C.~R., {Wyatt}, P., \& {Freedman}, W. 2006, in Bulletin of the
  American Astronomical Society, Vol.~38, Bulletin of the American Astronomical
  Society, 1026

\bibitem[{{Candia} {et~al.}(2003){Candia}, {Krisciunas}, {Suntzeff},
  {Gonz{\'a}lez}, {Espinoza}, {Leiton}, {Rest}, {Smith}, {Cuadra}, {Tavenner},
  {Logan}, {Snider}, {Thomas}, {West}, {Gonz{\'a}lez}, {Gonz{\'a}lez},
  {Phillips}, {Hastings}, \& {McMillan}}]{2003PASP..115..277C}
{Candia}, P., et al. 2003, \pasp, 115, 277

\bibitem[{{Cardelli} {et~al.}(1989){Cardelli}, {Clayton}, \&
  {Mathis}}]{1989ApJ...345..245C}
{Cardelli}, J.~A., {Clayton}, G.~C., \& {Mathis}, J.~S. 1989, \apj, 345, 245

\bibitem[{{de Jong}(2006)}]{2006ISR03}
{de Jong}, R. 2006, Instrument Science Report NICMOS 2006-003, (Baltimore, MD:
  STScI)

\bibitem[{{Shaw} \& {de Jong}(2008){Shaw},\& {de Jong}}]{2008ISR003}
         {Shaw}, B. \& {de Jong}, R. S. 2008, Instrument Science Report NICMOS 2008-003 (Baltimore, MD: STScI) 

\bibitem[{{Ellis} {et~al.}(2008){Ellis}, {Sullivan}, {Nugent}, {Howell},
  {Gal-Yam}, {Astier}, {Balam}, {Balland}, {Basa}, {Carlberg}, {Conley},
  {Fouchez}, {Guy}, {Hardin}, {Hook}, {Pain}, {Perrett}, {Pritchet}, \&
  {Regnault}}]{2008ApJ...674...51E}
{Ellis}, R.~S., et al. 2008, \apj, 674, 51

\bibitem[{{Fadeyev} {et~al.}(2006){Fadeyev}, {Aldering}, \&
  {Perlmutter}}]{2006PASP..118..907F}
{Fadeyev}, V., {Aldering}, G., \& {Perlmutter}, S. 2006, \pasp, 118, 907

\bibitem[{{Foley} {et~al.}(2008){Foley}, {Filippenko}, {Aguilera}, {Becker},
  {Blondin}, {Challis}, {Clocchiatti}, {Covarrubias}, {Davis}, {Garnavich},
  {Jha}, {Kirshner}, {Krisciunas}, {Leibundgut}, {Li}, {Matheson}, {Miceli},
  {Miknaitis}, {Pignata}, {Rest}, {Riess}, {Schmidt}, {Smith}, {Sollerman},
  {Spyromilio}, {Stubbs}, {Suntzeff}, {Tonry}, {Wood-Vasey}, \&
  {Zenteno}}]{2008ApJ...684...68F}
{Foley}, R.~J., et al. 2008, \apj, 684, 68

\bibitem[{{Garavini} {et~al.}(2004){Garavini}, {Folatelli}, {Goobar}, {Nobili},
  {Aldering}, {Amadon}, {Amanullah}, {Astier}, {Balland}, {Blanc}, {Burns},
  {Conley}, {Dahl{\'e}n}, {Deustua}, {Ellis}, {Fabbro}, {Fan}, {Frye}, {Gates},
  {Gibbons}, {Goldhaber}, {Goldman}, {Groom}, {Haissinski}, {Hardin}, {Hook},
  {Howell}, {Kasen}, {Kent}, {Kim}, {Knop}, {Lee}, {Lidman}, {Mendez},
  {Miller}, {Moniez}, {Mour{\~a}o}, {Newberg}, {Nugent}, {Pain}, {Perdereau},
  {Perlmutter}, {Prasad}, {Quimby}, {Raux}, {Regnault}, {Rich}, {Richards},
  {Ruiz-Lapuente}, {Sainton}, {Schaefer}, {Schahmaneche}, {Smith}, {Spadafora},
  {Stanishev}, {Walton}, {Wang}, \& {Wood-Vasey}}]{2004AJ....128..387G}
{Garavini}, G., et al. 2004, \aj, 128, 387

\bibitem[{{Garavini} {et~al.}(2007){Garavini}, {Folatelli}, {Nobili},
  {Aldering}, {Amanullah}, {Antilogus}, {Astier}, {Blanc}, {Bronder}, {Burns},
  {Conley}, {Deustua}, {Doi}, {Fabbro}, {Fadeyev}, {Gibbons}, {Goldhaber},
  {Goobar}, {Groom}, {Hook}, {Howell}, {Kashikawa}, {Kim}, {Kowalski},
  {Kuznetsova}, {Lee}, {Lidman}, {Mendez}, {Morokuma}, {Motohara}, {Nugent},
  {Pain}, {Perlmutter}, {Quimby}, {Raux}, {Regnault}, {Ruiz-Lapuente},
  {Sainton}, {Schahmaneche}, {Smith}, {Spadafora}, {Stanishev}, {Thomas},
  {Walton}, {Wang}, {Wood-Vasey}, \& {Yasuda}}]{2007A&A...470..411G}
{Garavini}, G., et al. 2007, \aap, 470, 411

\bibitem[{{Garnavich} {et~al.}(1998){Garnavich}, {Jha}, {Challis},
  {Clocchiatti}, {Diercks}, {Filippenko}, {Gilliland}, {Hogan}, {Kirshner},
  {Leibundgut}, {Phillips}, {Reiss}, {Riess}, {Schmidt}, {Schommer}, {Smith},
  {Spyromilio}, {Stubbs}, {Suntzeff}, {Tonry}, \&
  {Carroll}}]{1998ApJ...509...74G}
{Garnavich}, P.~M., et al. 1998, \apj, 509, 74

\bibitem[{{Goldhaber} {et~al.}(2001){Goldhaber}, {Groom}, {Kim}, {Aldering},
  {Astier}, {Conley}, {Deustua}, {Ellis}, {Fabbro}, {Fruchter}, {Goobar},
  {Hook}, {Irwin}, {Kim}, {Knop}, {Lidman}, {McMahon}, {Nugent}, {Pain},
  {Panagia}, {Pennypacker}, {Perlmutter}, {Ruiz-Lapuente}, {Schaefer},
  {Walton}, \& {York}}]{2001ApJ...558..359G}
{Goldhaber}, G., et al. \apj, 558, 359

\bibitem[{{Goobar}(2008)}]{2008ApJ...686L.103G}
{Goobar}, A. 2008, \apjl, 686, L103

\bibitem[{{Goobar} {et~al.}(2002){Goobar}, {Bergstr{\"o}m}, \&
  {M{\"o}rtsell}}]{2002A&A...384....1G}
{Goobar}, A., {Bergstr{\"o}m}, L., \& {M{\"o}rtsell}, E. 2002, \aap, 384, 1

\bibitem[{{Hoeflich} {et~al.}(1998){Hoeflich}, {Wheeler}, \&
  {Thielemann}}]{1998ApJ...495..617H}
{Hoeflich}, P., {Wheeler}, J.~C., \& {Thielemann}, F.~K. 1998, \apj, 495, 617

\bibitem[{{Howell} {et~al.}(2007){Howell}, {Sullivan}, {Conley}, \&
  {Carlberg}}]{2007ApJ...667L..37H}
{Howell}, D.~A., {Sullivan}, M., {Conley}, A., \& {Carlberg}, R. 2007, \apjl,
  667, L37

\bibitem[{{Howell} {et~al.}(2006){Howell}, {Sullivan}, {Nugent}, {Ellis},
  {Conley}, {Le Borgne}, {Carlberg}, {Guy}, {Balam}, {Basa}, {Fouchez}, {Hook},
  {Hsiao}, {Neill}, {Pain}, {Perrett}, \& {Pritchet}}]{2006Natur.443..308H}
{Howell}, D.~A., et al. 2006, \nat, 443, 308

\bibitem[{{Howell} {et~al.}(2005){Howell}, {Sullivan}, {Perrett}, {Bronder},
  {Hook}, {Astier}, {Aubourg}, {Balam}, {Basa}, {Carlberg}, {Fabbro},
  {Fouchez}, {Guy}, {Lafoux}, {Neill}, {Pain}, {Palanque-Delabrouille},
  {Pritchet}, {Regnault}, {Rich}, {Taillet}, {Knop}, {McMahon}, {Perlmutter},
  \& {Walton}}]{2005ApJ...634.1190H}
{Howell}, D.~A., et al. 2005, \apj, 634, 1190

\bibitem[{{Hsiao} {et~al.}(2007){Hsiao}, {Conley}, {Howell}, {Sullivan},
  {Pritchet}, {Carlberg}, {Nugent}, \& {Phillips}}]{2007ApJ...663.1187H}
{Hsiao}, E.~Y., et al. 2007, \apj,
  663, 1187

\bibitem[{{Kim} {et~al.}(1996){Kim}, {Goobar}, \&
  {Perlmutter}}]{1996PASP..108..190K}
{Kim}, A., {Goobar}, A., \& {Perlmutter}, S. 1996, \pasp, 108, 190

\bibitem[{{Knop} {et~al.}(2003){Knop}, {Aldering}, {Amanullah}, {Astier},
  {Blanc}, {Burns}, {Conley}, {Deustua}, {Doi}, {Ellis}, {Fabbro}, {Folatelli},
  {Fruchter}, {Garavini}, {Garmond}, {Garton}, {Gibbons}, {Goldhaber},
  {Goobar}, {Groom}, {Hardin}, {Hook}, {Howell}, {Kim}, {Lee}, {Lidman},
  {Mendez}, {Nobili}, {Nugent}, {Pain}, {Panagia}, {Pennypacker}, {Perlmutter},
  {Quimby}, {Raux}, {Regnault}, {Ruiz-Lapuente}, {Sainton}, {Schaefer},
  {Schahmaneche}, {Smith}, {Spadafora}, {Stanishev}, {Sullivan}, {Walton},
  {Wang}, {Wood-Vasey}, \& {Yasuda}}]{2003ApJ...598..102K}
{Knop}, R.~A., et al. 2003, \apj, 598, 102

\bibitem[{{Kowalski} {et~al.}(2008){Kowalski}, {Rubin}, {Aldering},
  {Agostinho}, {Amadon}, {Amanullah}, {Balland}, {Barbary}, {Blanc}, {Challis},
  {Conley}, {Connolly}, {Covarrubias}, {Dawson}, {Deustua}, {Ellis}, {Fabbro},
  {Fadeyev}, {Fan}, {Farris}, {Folatelli}, {Frye}, {Garavini}, {Gates},
  {Germany}, {Goldhaber}, {Goldman}, {Goobar}, {Groom}, {Haissinski}, {Hardin},
  {Hook}, {Kent}, {Kim}, {Knop}, {Lidman}, {Linder}, {Mendez}, {Meyers},
  {Miller}, {Moniez}, {Mourao}, {Newberg}, {Nobili}, {Nugent}, {Pain},
  {Perdereau}, {Perlmutter}, {Phillips}, {Prasad}, {Quimby}, {Regnault},
  {Rich}, {Rubenstein}, {Ruiz-Lapuente}, {Santos}, {Schaefer}, {Schommer},
  {Smith}, {Soderberg}, {Spadafora}, {Strolger}, {Strovink}, {Suntzeff},
  {Suzuki}, {Thomas}, {Walton}, {Wang}, {Wood-Vasey}, \&
  {Yun}}]{2008ApJ...686..749K}
{Kowalski}, M., et al. 2008, ApJ, 686, 749

\bibitem[{{Krist} \& Hook(2004)}]{2004krist.book}
{Krist}, J., \& Hook, R. 2004, The Tiny Tim User's Guide. Version 6.3
  (Baltimore, MD: STScI)

\bibitem[{{Lentz} {et~al.}(2000){Lentz}, {Baron}, {Branch}, {Hauschildt}, \&
  {Nugent}}]{2000ApJ...530..966L}
{Lentz}, E.~J., et al. 2000, \apj, 530, 966

\bibitem[{{Li} {et~al.}(2001{\natexlab{a}}){Li}, {Filippenko}, {Gates},
  {Chornock}, {Gal-Yam}, {Ofek}, {Leonard}, {Modjaz}, {Rich}, {Riess}, \&
  {Treffers}}]{2001PASP..113.1178L}
{Li}, W., et al. \pasp, 113, 1178

\bibitem[{{Li} {et~al.}(2001{\natexlab{b}}){Li}, {Filippenko}, {Treffers},
  {Riess}, {Hu}, \& {Qiu}}]{2001ApJ...546..734L}
{Li}, W., et al. 2001{\natexlab{b}}, \apj, 546, 734

\bibitem[{{Matheson} {et~al.}(2005){Matheson}, {Blondin}, {Foley}, {Chornock},
  {Filippenko}, {Leibundgut}, {Smith}, {Sollerman}, {Spyromilio}, {Kirshner},
  {Clocchiatti}, {Aguilera}, {Barris}, {Becker}, {Challis}, {Covarrubias},
  {Garnavich}, {Hicken}, {Jha}, {Krisciunas}, {Li}, {Miceli}, {Miknaitis},
  {Prieto}, {Rest}, {Riess}, {Salvo}, {Schmidt}, {Stubbs}, {Suntzeff}, \&
  {Tonry}}]{2005AJ....129.2352M}
{Matheson}, T. et al. 2005, \aj, 129,
  2352

\bibitem[{{Nobili} {et~al.}(2005){Nobili}, {Amanullah}, {Garavini}, {Goobar},
  {Lidman}, {Stanishev}, {Aldering}, {Antilogus}, {Astier}, {Burns}, {Conley},
  {Deustua}, {Ellis}, {Fabbro}, {Fadeyev}, {Folatelli}, {Gibbons}, {Goldhaber},
  {Groom}, {Hook}, {Howell}, {Kim}, {Knop}, {Nugent}, {Pain}, {Perlmutter},
  {Quimby}, {Raux}, {Regnault}, {Ruiz-Lapuente}, {Sainton}, {Schahmaneche},
  {Smith}, {Spadafora}, {Thomas}, {Wang}, \& {The Supernova Cosmology
  Project}}]{2005A&A...437..789N}
{Nobili}, S., et al. 2005, \aap, 437, 789

\bibitem[{{Nobili} \& {Goobar}(2008)}]{2008A&A...487...19N}
{Nobili}, S., \& {Goobar}, A. 2008, \aap, 487, 19

\bibitem[{{Nordin} {et~al.}(2008){Nordin}, {Goobar}, \&
  {J{\"o}nsson}}]{2008JCAP...02..008N}
{Nordin}, J., {Goobar}, A., \& {J{\"o}nsson}, J. 2008, Journal of Cosmology and
  Astro-Particle Physics, 2, 8

\bibitem[{{Nugent} {et~al.}(2002){Nugent}, {Kim}, \&
  {Perlmutter}}]{2002PASP..114..803N}
{Nugent}, P., {Kim}, A., \& {Perlmutter}, S. 2002, \pasp, 114, 803

\bibitem[{{Perlmutter} {et~al.}(1998){Perlmutter}, {Aldering}, {della Valle},
  {Deustua}, {Ellis}, {Fabbro}, {Fruchter}, {Goldhaber}, {Groom}, {Hook},
  {Kim}, {Kim}, {Knop}, {Lidman}, {McMahon}, {Nugent}, {Pain}, {Panagia},
  {Pennypacker}, {Ruiz-Lapuente}, {Schaefer}, \&
  {Walton}}]{1998Natur.391...51P}
{Perlmutter}, S., et al. 1998, \nat, 391, 51

\bibitem[{{Perlmutter} {et~al.}(1999){Perlmutter}, {Aldering}, {Goldhaber},
  {Knop}, {Nugent}, {Castro}, {Deustua}, {Fabbro}, {Goobar}, {Groom}, {Hook},
  {Kim}, {Kim}, {Lee}, {Nunes}, {Pain}, {Pennypacker}, {Quimby}, {Lidman},
  {Ellis}, {Irwin}, {McMahon}, {Ruiz-Lapuente}, {Walton}, {Schaefer}, {Boyle},
  {Filippenko}, {Matheson}, {Fruchter}, {Panagia}, {Newberg}, {Couch}, \& {The
  Supernova Cosmology Project}}]{1999ApJ...517..565P}
{Perlmutter}, S., et al. 1999, \apj, 517, 565

\bibitem[{{Perlmutter} {et~al.}(1997){Perlmutter}, {Gabi}, {Goldhaber},
  {Goobar}, {Groom}, {Hook}, {Kim}, {Kim}, {Lee}, {Pain}, {Pennypacker},
  {Small}, {Ellis}, {McMahon}, {Boyle}, {Bunclark}, {Carter}, {Irwin},
  {Glazebrook}, {Newberg}, {Filippenko}, {Matheson}, {Dopita}, {Couch}, \& {The
  Supernova Cosmology Project}}]{1997ApJ...483..565P}
{Perlmutter}, S., et al. 1997,
  \apj, 483, 565

\bibitem[{{Riess} {et~al.}(1998){Riess}, {Filippenko}, {Challis},
  {Clocchiatti}, {Diercks}, {Garnavich}, {Gilliland}, {Hogan}, {Jha},
  {Kirshner}, {Leibundgut}, {Phillips}, {Reiss}, {Schmidt}, {Schommer},
  {Smith}, {Spyromilio}, {Stubbs}, {Suntzeff}, \&
  {Tonry}}]{1998AJ....116.1009R}
{Riess}, A.~G., et al. 1998, \aj, 116, 1009

\bibitem[{{Riess} {et~al.}(2007){Riess}, {Strolger}, {Casertano}, {Ferguson},
  {Mobasher}, {Gold}, {Challis}, {Filippenko}, {Jha}, {Li}, {Tonry}, {Foley},
  {Kirshner}, {Dickinson}, {MacDonald}, {Eisenstein}, {Livio}, {Younger}, {Xu},
  {Dahl{\'e}n}, \& {Stern}}]{2007ApJ...659...98R}
{Riess}, A.~G., et al. 2007, \apj, 659, 98

\bibitem[{{Sauer} {et~al.}(2008){Sauer}, {Mazzali}, {Blondin}, {Filippenko},
  {Benetti}, {Stehle}, {Challis}, {Kirshner}, \& {Li}}]{2008MNRAS.391.1605S}
{Sauer}, D.~N., et al.2008, \mnras, 391, 1605 

\bibitem[{{Schlegel} {et~al.}(1998){Schlegel}, {Finkbeiner}, \&
  {Davis}}]{1998ApJ...500..525S}
{Schlegel}, D.~J., {Finkbeiner}, D.~P., \& {Davis}, M. 1998, \apj, 500, 525

\bibitem[{{Schmidt} {et~al.}(1998){Schmidt}, {Suntzeff}, {Phillips},
  {Schommer}, {Clocchiatti}, {Kirshner}, {Garnavich}, {Challis}, {Leibundgut},
  {Spyromilio}, {Riess}, {Filippenko}, {Hamuy}, {Smith}, {Hogan}, {Stubbs},
  {Diercks}, {Reiss}, {Gilliland}, {Tonry}, {Maza}, {Dressler}, {Walsh}, \&
  {Ciardullo}}]{1998ApJ...507...46S}
{Schmidt}, B.~P., et al. 1998, \apj, 507, 46

\bibitem[{{Tonry} {et~al.}(2003){Tonry}, {Schmidt}, {Barris}, {Candia},
  {Challis}, {Clocchiatti}, {Coil}, {Filippenko}, {Garnavich}, {Hogan},
  {Holland}, {Jha}, {Kirshner}, {Krisciunas}, {Leibundgut}, {Li}, {Matheson},
  {Phillips}, {Riess}, {Schommer}, {Smith}, {Sollerman}, {Spyromilio},
  {Stubbs}, \& {Suntzeff}}]{2003ApJ...594....1T}
{Tonry}, J.~L., et al. 2003, \apj, 594, 1

\bibitem[{{Wang}(2005)}]{2005ApJ...635L..33W}
{Wang}, L. 2005, \apjl, 635, L33

\bibitem[{{Wood-Vasey} {et~al.}(2008{\natexlab{a}}){Wood-Vasey}, {Friedman},
  {Bloom}, {Hicken}, {Modjaz}, {Kirshner}, {Starr}, {Blake}, {Falco},
  {Szentgyorgyi}, {Challis}, {Blondin}, \& {Rest}}]{2008ApJ...689..377W}
{Wood-Vasey}, W.~M., et al. 2008, \apj, 689, 377

\bibitem[{{Wood-Vasey} {et~al.}(2007{\natexlab{b}}){Wood-Vasey}, {Miknaitis},
  {Stubbs}, {Jha}, {Riess}, {Garnavich}, {Kirshner}, {Aguilera}, {Becker},
  {Blackman}, {Blondin}, {Challis}, {Clocchiatti}, {Conley}, {Covarrubias},
  {Davis}, {Filippenko}, {Foley}, {Garg}, {Hicken}, {Krisciunas}, {Leibundgut},
  {Li}, {Matheson}, {Miceli}, {Narayan}, {Pignata}, {Prieto}, {Rest}, {Salvo},
  {Schmidt}, {Smith}, {Sollerman}, {Spyromilio}, {Tonry}, {Suntzeff}, \&
  {Zenteno}}]{2007ApJ...666..694W}
{Wood-Vasey}, W.~M., et al. 2007, \apj, 666, 694

\end{thebibliography}

\end{document}